\documentclass[authoryear,1p]{elsarticle}
\usepackage{graphicx}
\usepackage{array}
\usepackage[table]{xcolor}
\usepackage{multirow}
\usepackage{amssymb}
\usepackage{booktabs}
\usepackage{empheq}
\usepackage{natbib}
\usepackage[toc]{appendix}
\usepackage{xcolor}
\usepackage{eurosym}
\usepackage{url}
\usepackage{float}
\usepackage[T1]{fontenc}
\usepackage[utf8]{inputenc}
\usepackage{lmodern}
\usepackage{tikz}
\def\checkmark{\tikz\fill[scale=0.4](0,.35) -- (.25,0) -- (1,.7) -- (.25,.15) -- cycle;}
\usepackage{nomencl}
\usepackage{framed}
\usepackage{multicol}
\makenomenclature

\usepackage{mathtools}

\newdefinition{rmk}{Remark}
\newtheorem{proof}{Proof}
\newtheorem{ass}{Assumption}
\newtheorem{prop}{Proposition}
\usepackage[english]{babel}
\usepackage{hyperref}
\newcommand{\sref}[2]{\hyperref[#2]{#1 \ref*{#2}}}

\let\oldref\ref
\renewcommand{\ref}[1]{(\oldref{#1})}
\usepackage[nameinlink, capitalise, noabbrev]{cleveref}
\hypersetup{
	colorlinks,
	citecolor=blue,
	filecolor=blue,
	linkcolor=blue,
	urlcolor=blue}
\crefname{prop}{Proposition}{Propositions}
\crefname{thm}{Theorem}{Theorems}
\crefname{cor}{Corollary}{Corollaries}
\crefname{ass}{Assumption}{Assumptions}
\makenomenclature
\usepackage{etoolbox}
\renewcommand\nomgroup[1]{%
	\item[\bfseries
	\ifstrequal{#1}{A}{Sets}{%
	\ifstrequal{#1}{F}{Futures Market Parameters}{%
	\ifstrequal{#1}{V}{Variables}{%
	\ifstrequal{#1}{M}{Spot Market Parameters}{}}}}%
	]}
	\makeatletter
\def\bs{\expandafter\@gobble\string\\}
\def\lb{\expandafter\@gobble\string\{}
\def\rb{\expandafter\@gobble\string\}}
\def\@pdfauthor{C.V.Radhakrishnan}
\def\@pdftitle{elsarticle.cls -- A documentation}
\def\@pdfsubject{Document formatting with elsarticle.cls}
\def\@pdfkeywords{LaTeX, Elsevier Ltd, document class}

\DeclareRobustCommand{\LaTeX}{L\kern-.26em%
        {\sbox\z@ T%
         \vbox to\ht\z@{\hbox{\check@mathfonts
           \fontsize\sf@size\z@
           \math@fontsfalse\selectfont
          A\,}%
         \vss}%
        }%
     \kern-.15em%
    \TeX}

\journal{Energy Economics}

\begin{document}
	\begin{frontmatter}
		\title{Contracts in Electricity Markets under EU ETS: A Stochastic Programming Approach}
		\author[1]{Arega Getaneh Abate}
		\ead{a.abate004@unibs.it}
		\author[1]{Rossana Riccardi\corref{cor1}}
		\ead{rossana.riccardi@unibs.it}
		\author[2,3]{Carlos Ruiz}
		\ead{caruizm@est-econ.uc3m.es}
		\cortext[cor1]{Corresponding author}
		\address[1]{Department of Economics and Management,
			University of Brescia, 74/B Brescia, Italy}
		\address[2]{Department of Statistics, University Carlos III de Madrid, Avda. de la Universidad 30, 28911-Leganés, Spain.}
		\address[3]{UC3M-BS Institute for Financial Big Data (FiBiD), Universidad Carlos III de Madrid, 28903, Getafe, Madrid, Spain.}
		
		\begin{abstract}
The European Union Emission Trading Scheme (EU ETS) is a cornerstone of the EU's strategy to fight climate change and an important device for plummeting greenhouse gas (GHG) emissions in an economically efficient manner. The power industry has switched to an auction-based allocation system at the onset of Phase III of the EU ETS to bring economic efficiency by negating windfall profits that have been resulted from grandfathered allocation of allowances in the previous phases. In this work, we analyze and simulate the interaction of oligopolistic generators in an electricity market with a game-theoretical framework where the electricity and the emissions markets interact in a two-stage electricity market. For analytical simplicity, we assume a single futures market where the electricity is committed at the futures price, and the emissions allowance is contracted in advance, prior to a spot market where the energy and allowances delivery takes place. Moreover, a coherent risk measure is applied (Conditional Value at Risk) to model both risk averse and risk neutral generators and a two-stage stochastic optimization setting is introduced to deal with the uncertainty of renewable capacity, demand, generation, and emission costs. The performance of the proposed equilibrium model and its main properties are examined through realistic numerical simulations. Our results show that renewable generators are surging and substituting conventional generators without compromising social welfare. Hence, both renewable deployment and emission allowance auctioning are effectively reducing GHG emissions and promoting low-carbon economic path.
		\end{abstract}
	\end{frontmatter}
	\section{Introduction}\label{S_1}
Global warming has become one of the crucial environmental issues that has derived the attention of researchers and politicians for the last three decades. To curb the serious damage of global warming, the United Nations Framework Convention on Climate Change (UNFCCC) established the Kyoto Protocol in 1997 and entered into force on 16 February 2005. The EU had committed to reducing GHG emissions by 8\%, compared to the 1990 level, by the years 2008-2012 through a trading scheme in sectors with an installed capacity of more than 20 MW \citep{thema2013impact}. The EU ETS is the largest single market for CO$_2$ emission allowances, accounting for approximately 84\% of the global carbon market value \citep{action2013eu} and aimed to aid the EU's fight against the global warming.

GHG emissions, especially carbon dioxide (CO$_2$), and climate change are closely related to economic growth and social development. One of the main mechanisms of GHG reduction dictated by the protocol was to trade human-related emission allowances or permits, primarily CO$_2$, in organized financial markets\footnote{Since the electricity industry is the largest contributor of GHG, it should be considered in any global emissions combating policy initiative to reach decarbonized economy.}. Thus, several national and regional emission markets have been established where a variety of specialized financial instruments are traded \citep{lau2012global,moore2016will}.

Being the largest emissions trading system, the EU ETS covers more than 11,000 power factories and other installations in 31 countries, and flights between airports of participating countries which cover around 45\% of the EU's GHG emissions \citep{action2013eu,Commission_revision_2020}. According to the European Commission, to achieve the EU's overall GHG emissions reduction target for 2030, the sectors covered by the EU ETS must reduce their emissions by 43\% compared to 2005 levels. This target is supposed to be achieved by putting a price on carbon and thereby giving a financial value to each ton of emissions saved. Therefore, CO$_2$ is priced and leads electricity generators to invest in low-carbon technologies which, in turn discourage carbon-intensive generators by increasing their production cost\footnote{From its commencement, the EU ETS has progressed four distinct trading compliance phases to date: Phase I 2005-2007, Phase II 2008-2012, Phase III 2013-2020, and Phase IV 2021-2030.}.  
    
The details of the ETS application have been an important issue since the beginning. It commenced as cap-and-trade\footnote{Cap-and-trade is the most discussed scheme to control CO$_2$ emissions in the European Union trading system.} principle where the overall volume of GHG that can be emitted by the power plants and other companies covered by the system was subject to a cap set at the EU level. The emission cap was allocated for free of charge to power producers in the form of carbon permits, the so-called European emission allowances (EUAs), where one EUA gives the owner the right to release one ton of CO$_2$ in the atmosphere\footnote{Within the cap, companies receive or buy emission allowances which they can trade, if they wish to do so in the later phases.}. If electricity generators have emission discrepancy, they should turn to the EU ETS market to sell/buy any surplus/shortage of permits. For instance, if they emit less than their cap, they can sell the surplus of permits. Otherwise, they pay the penalties if they emit above the cap. However, studies from the earlier phases show that allocating emissions based on the cap-and-trade strategy is not as efficient as expected \citep{veith2009capital,ellerman2016european}.
    
    The main issue concerning the operational allocation method based on grandfathering emitters by assigning emission allowances free of charge (based on historical emissions) had enabled compliant electricity producers to generate large carbon rents or windfall profits \citep{veith2009capital,ellerman2016european}. Moreover, the allocation of allowances to member states based on grandfathering is criticized for the alleged competitive distortions \citep{ellerman2016european}. One of the distortions was that companies generated windfall profits by reporting costs above reality. Windfall profits stifle investments in low-carbon technologies and in the worst-case encourage investment in carbon-intensive installations. As a solution, with Directive 2009/29/EC, a significant shift towards an exclusively auction-based system for allowances to the power sector was instituted at the onset of Phase III. After Phase III, power producers in most member states (except new member states who are permitted to obtain EUA free of charge) purchase allowances in accordance with the polluter pays principle \citep{ hobbie2019windfall}. The reform aims to negate windfall profits being earned by producers in the power sector by forcing electricity producers to assume material costs for the permits obtained. Auctioning addresses both criticisms of the previous grandfathering allocations in one fell swoop. First, windfall profits are eliminated through the auctioning of allowances. Second, the possibility of competitive distortions within the single EU market is solved by the force of demand and supply. Since the auction-based methods are increasingly becoming the preferred allocation mechanism by policy-makers\footnote{The auction markets for carbon allowances can effectively avoid the intrinsic shortcomings of the centralized allocation methods, political misallocation, and regulatory distortion \citep{tang2017carbon}.}, we assume in our model that the allowances are auctioned instead of allocating for free (see for example \cite{rocha2015impact} and \cite{tang2017carbon} for a discussion of this issue).
    
    Another instrument to combat the GHG effect is the expansion of renewable energy sources. The deployment of intermittent capacities, such as wind and solar, are effective instruments that can be done at a lower cost in the electricity sector with less structural changes than in other sectors. However, there are debates that policy combinations of renewable deployment and devising EU ETS to combat GHG have negatively interacted as the EU ETS has a dampening effect on the CO$_2$ price \citep{ del2017does}. The existence of such different policy goals and market failures are not panaceas and bring problems on their own. Rather there are arguments that the economic theory supports the combination of ETS and renewable energy targets by mitigating policy trade-offs through appropriate coordination and/or instrument choice and design, as recommended by \cite{ del2017does} and \cite{munoz2017does}.
    
    In this paper, we analyze the interaction of electricity generators in an oligopolistic market with a game-theoretical model where the electricity market and the emissions market interact in a two-stage framework. For analytical simplicity, we assume a single futures electricity market where a quantity is committed at the futures price. Similarly, the emissions allowance is contracted in advance prior to a spot market where the electricity and allowances delivery takes place. We represent these market interactions with a game-theoretical equilibrium model. Both risk neutral and risk averse generators are taken into consideration with different hedging strategies. In order to achieve that, a nonlinear programming problem is solved after reformulating the CVaR optimization problem for each agent via its equivalent Karush-Kuhn-Tucker (KKT) conditions. Finally, the model is tested numerically with calibrated data. Our findings show that both RES and CO$_2$ parameters are effective in plummeting GHG by suppressing conventional generators and helping the low-carbon transition without compromising social welfare.
    
    The paper is organized as follows. \cref{S_2} reviews the literature on electricity, emission allowance, and GHG reduction. \cref{S_3} presents the game-theoretical stochastic programming model formulation and the approaches to solve it with futures emission allowances and RES parameters. \cref{S_4} tests the model with calibrated data and presents relevant numerical results. Finally, we provide concluding remarks in \cref{S_5}.
	
\section{Literature Review and Contributions}\label{S_2}
		\subsection{Literature Review}
	\subsubsection{EU ETS and Electricity Market}
	Since the 1990s, there has been a pouring of research efforts to assess and to combat the effects of GHG emissions through different strategies, like cap-and-trade programs in Europe.
    The European Union has created a single European electricity market, operating under ETS in the first half of the 2000s. That coincided with electricity market liberalization. Market liberalization and increasing transmission capacity between regional electricity grids have been served as the key means for the promotion of internal electricity market creation and competition \citep{chevallier2011model,aatola2013impact}. This further promotes the creation and enhancement of the regulation between European countries. One of the economic activities in the regulation is the carbon market that has got the attention of countries around the world. The carbon market is considered an economically efficient mechanism to reduce the cost of emission abatement by institutional innovation and trading in financial markets \citep{cui2014will}. Recent studies reveal that integrating GHG combating with emission trading is quite significant to achieve economic and environmental objectives. EU ETS is an incentive-based and market-driven instrument purported to achieve environmental objectives by allocating and auctioning GHG emission allowances. In other words, the EU ETS is taken as an important pathway of development for the EU to respond to the global climate change calls by achieving simultaneously green and low-carbon economic development \citep{naso2017porter,wang2019can}. Because economic activities foster higher demand for industrial production, companies (such as electricity generators) falling under the regulation of the EU ETS, need to produce more in order to meet their customers' demand. This, in turn yields large demand for CO$_2$ allowances (as they emit more CO$_2$) to cover emissions shortage which ultimately leads to the CO$_2$ price increase \citep{ellerman2008over}.
    
    The CO$_2$ price is determined by the interplay between supply and demand. In response to low emission prices, recently, the EU is deciding to harness the emission allowance supply through backloading\footnote{Backloading is a decision to postpone auctions of a certain number of allowances. This measure aimed at re-balancing supply and demand of allowances in the short term to improve the overall functioning of the EU ETS.} and the Market Stability Reserve (MSR)\footnote{The MSR is a carbon market reform aimed at providing price stability for installations covered under the EU ETS scheme and spurring low-carbon investments by increasing resilience to demand-supply imbalances. The MSR reduces yearly allocations depending on the size of the surplus. Later, it injects stored allowances back into the market once a certain scarcity threshold is reached. Hence, it shifts the auction date of allowances into the future \citep{ perino2017eu}. The MSR is a mechanism established by the EU in 2015 to reduce the surplus of emission allowances in the carbon market and to improve the EU ETS's resilience to future shocks} which are effected in 2019 so that both measures are directed at increasing prices \citep{salant2016ails}. Therefore, the EU electricity market is an area within which the EU ETS and energy efficiency instruments interact with \citep{ thema2013impact}. After the onset of Phase III of the EU ETS, the electricity generating sector and energy-intensive industries are obliged to buy/sell from the auction market (as a function of market outcomes) that replaces the fixed cap with their emission permit surplus/shortage.  
    \subsubsection{Emission Allowances Models in Electricity Market}
    The linkage between carbon futures markets with the energy market and the effectiveness of the carbon futures markets to combat GHG are two strands of studies relevant to this topic. This dynamic development of financial markets and carbon futures has attracted numerous researchers and practitioners with quantitative models \citep{wu2011dynamic}. More often, the focus on the interaction between power systems and carbon policies is considered from the power market and emission reduction perspectives. In this regard, \cite{rocha2015impact} analyze the interactions between a cap-and-trade program, investment decisions, and electricity markets for an electricity network in a restructured market environment by considering the CO$_2$ emissions. Another simulation model is done by \cite{ruth2008economic} to examine the energy and economic implications whereby they conclude modest emissions reductions, despite profit is reduced when conventional plants are retired. \cite{wang2020carbon} propose an agent-based approach with Q-learning algorithm to model\footnote{Q-learning algorithm is a model-free reinforcement learning algorithm to learn a policy telling an agent what action to take under what circumstances.} interactions between the carbon trading market where carbon emission is regulated from the electricity generators side. On the other hand, since the carbon market is risk prone, it attracts attention in the field of energy economics and finance. In the carbon market, the market risk comes not only from its internal volatility but also from multiple aspects (generators technologies, costs of production uncertainties, demand, and nonstorability of the energy generated) \citep{hintermann2010allowance,lin2019main} 
    
    In general, given that allowances trading has primarily been applied in the EU ETS, there are quite a lot of studies that analyze price behavior of tradable emission allowances from different perspectives. For instances, \cite{daskalakis2009modeling,seifert2008dynamic,paolella2006econometric} from econometrics and management perspectives, \cite{bohringer2005economic,daskalakis2009modeling,samadi2018analyzing} with price simulations against changes in market design parameters, \cite{galiana2010emission,el2019investigating} with deterministic Cournot-Nash electricity companies gaming strategies and emission permit, \cite{chevallier2010modelling,botterud2010relationship,boersen2014relationship} with futures/forward contracts and EUAs allowances, and earlier \cite{chuang2001game} using the competition of generators and Cournot theory of oligopoly, among others. Methodologically, there are estimation and optimization models. They have contributed to the development of the allowance market. However, there are still gaps in the literature that combines CO$_2$ allowances allocation through carbon market auction and RES penetration with stochastic programming\footnote{ Stochastic programming method associated with scenario generation approach which has been extensively deployed in problems which include uncertainties. In such programming approach, the uncertain terms are expressed as stochastic variables and represented with scenarios \citep{hajibandeh2017new}}.
    \subsubsection{Risk Aversion Models in Electricity Market}
Risk management is a crucial task that has been considered in decision science. There are numerous ways to include risk aversion in market equilibrium models. For example, in the economics literature, concave utility functions\footnote{That is, the utility functions can be specified as exponential functions (exhibiting constant absolute risk aversion, CARA) and isoelastic functions (exhibiting constant relative risk aversion, CRRA) \citep{eeckhoudt2011economic}.} are popular as they can be used to convert monetary costs/profits into utilities, whose expected value is then optimized instead of the original objective \citep{ fishburn1970utility}. In such cases, the non-linearity nature of the functions makes them challenging to be included in large-scale market equilibrium models. In addition, if the distribution of the possible outcomes is normal, the exponential function can be written as a linear combination of expected outcomes and the standard deviation of the outcome distribution. This further complicates the model as the normality assumption is not realistic in the case of considering an small number of scenarios, which is usually the case due to computational complexity.
As a result, in economics and finance, there are sophisticated theoretical and quantitative tours-de-force risk measures such as variance, shortfall probability, expected shortage, VaR, CVaR and stochastic dominance, which can control the trade-off between expected profit and the variability of profit. 
    
Specifically, in financial mathematics literature, risk aversion is modeled by including VaR \citep{duffie1997overview}, or CVaR \citep{rockafellar2000optimization} in the decision maker’s objective, or constraints. However, VaR has undesirable mathematical properties\footnote{According to \cite{uryasev2010var}, VaR has the following undesirable mathematical properties:
    \begin{enumerate}
\item It does not control scenarios exceeding VaR
\item It is a non-convex and discontinuous function of the confidence level for discrete distributions
\item VaR is difficult to control/optimize for non-normal distributions.
\end{enumerate} Whereas CVaR is continuous with respect to the confidence level and convex in decision variables.}. CVaR\footnote{CVaR is a coherent risk measure (i.e., it exhibits good properties such as translation invariance, subadditivity, positive homogeneity, and monotonicity) that was introduced by \cite{rockafellar2000optimization} as a technique for portfolio optimization which calculates VaR and optimizes CVaR simultaneously.}, on the contrary, gives the expected value over outcomes that are worse than the VaR.    

Our model applies CVaR risk aversion on profit optimization with financial derivatives in the electricity market and emission auctioning. We formulate a stochastic model that maximizes a weighted average of expected profits and their CVaR. In particular, we use stochastic programming  because: i) it provides a powerful framework to model and include parameters' uncertainty in an optimization problem, via a plausible set of scenarios; and ii) the CVaR can be easily incorporated to the model with a linear formulation (\citep{rockafellar2000optimization}).

The CVaR at $\alpha$ confidence level (CVaR$_{\alpha}$) can be defined as the expected value of the profit smaller than the ($1-\alpha$)-quantile of the profit distribution. 
CVaR has been widely used in various electricity power problems for different entities for different purposes \citep{artzner1999application,zhang2006competition,conejo2008optimal,shapiro2009time,morales2010short,philpott2013solving,del2017does}. For instance, \cite{ neuhoff2004insufficient} use for competitive markets where risks cannot be traded and coupled with risk averse generators and consumers. \cite{ehrenmann2011generation} show the generation cost increasing and market failure by applying stochastic equilibrium models with CVaR maximizing investors' return. In their models, they feature uncertain fuel costs, emissions reduction targets, and numbers of carbon allowances where risk averse investors build more open cycle gas turbines and less coal-fired generation capacity.
    
The novelty in our work is the analysis of the interaction between different financial derivatives (electricity and emission permits) and markets (futures and spot) from a game-theoretical perspective. Furthermore, we try to characterize the equilibrium between risk averse generators (CVaR) under different levels of competition in the market. Hence, our model combines different market designs and interactions between energy markets and financial derivatives. To this end, we develop a two-stage stochastic programming model which has a closed-form solution in the second stage and allows to be reformulated as a single nonlinear optimization problem. Subsequently, we examine strategic players reactions in both electricity and emission allowance markets through realistic numerical simulations.
    \subsection{Contributions of the Paper}
In this work, we develop market equilibrium models that combine emissions trading and RES deployment with oligopolistic risk neutral/averse players. The models are examined with trading data calibrated from European Energy Exchange (EEX) and the Spanish electricity market which implicate the main entities (generators, marketers, consumers, regulators) involved in power markets and climate change. The main contributions of this work are fivefold:
    \begin{enumerate}
    \item Modeling ad-hoc electricity market that involves renewable penetration and emission reduction simultaneously with a single futures market and a subsequent spot market.
    \item Developing and examining a two-stage stochastic programming model with risk neutral/averse generators by introducing coherent risk measurement (CVaR) and different levels of competition (Cournot and perfect competition).
    \item Deriving analytical expressions to characterize the equilibrium in both the futures and spot markets with emission futures and electricity futures, considering RES and CO$_2$ as parameters.
    \item Testing the equilibrium model with relevant and insightful data to highlight the impact of low-carbon energy expansion and combating emission via EUA auctioning instead of grandfathering (allocations in the previous stages of the regulations).
    \item Examining how to achieve the targets of plummeting GHG emissions in an economically efficient manner using both RES penetration and CO$_2$ auctioning in the derivatives market.
    \end{enumerate}
	\section{Problem Formulation}\label{S_3}
    In this section, we present the model formulation that entails two specific models. The benchmark is referred to as the general model (GM) which includes a futures contract for both emission trading and electricity market in a two-stage stochastic programming framework. The second model examines a single spot market model for comparison and completeness.
   \begin{itemize}
  	\item[] {\bf{Sets}}
  	\begin{itemize}
  		\item[] {$I$} \qquad {Set of conventional generators ranging from $i =1,..., \mid I \mid $}
  		\item[] {$J$} \qquad {Set of renewable resource generators ranging from $j=1,...,\mid J\mid$}
  		\item[] {$\Omega$} \qquad {Set of scenarios ranging from $\omega =1,...,\mid\Omega\mid$}
  	\end{itemize}
  	\item[] {\bf{Futures Market Parameters}}
  	\begin{itemize}
  		\item[] {$\beta^F$}\qquad\qquad{Electricity price demand slope in the futures market}
  		\item[] {$\gamma^{F}$}\qquad\qquad{Electricity price demand intercept in the futures market}
  		\item[] {$P^{F_{CO_2}}$}\qquad\ {Price of emission allowances [\euro/ton CO$_2$] in the futures market} 
  	\end{itemize}
  	\item[] {\bf{Spot Market Parameters}}
  	\begin{itemize}
  		\item[]{$\beta_{\omega}^{S}$}\qquad\ {Electricity price demand slope in the spot market}
  		\item[] {$\hat{\gamma_{\omega}}^S$}\qquad{Electricity price demand intercept in the spot market}
  		\item[] {$\hat{\eta_{i\omega}}$}\qquad\ {Emissions intensity factor for generator $i$ [ton CO$_2$/MWh]}
  		\item[] {$Q_{j\omega}$}\qquad{Renewable energy total production for generator $j$ [MWh]}
  	\end{itemize}

  	\item[] {\bf{Variables}}
  	\begin{itemize}
  		\item[] {$P_{\omega}^{S_{CO_2}}$}\qquad {Price of emission allowances [\euro/ton CO$_2$] in the spot market}
  		\item[] {$P^F$}\qquad\quad \ {Electricity price [\euro/MWh] in the futures market}
  		\item[] {$P_{\omega}^S$}\qquad\quad \  {Electricity price [\euro/MWh] in the spot market}
  		\item[] {$q_i^F$}\qquad\quad \  \ {Futures market electricity quantity [MWh]}
  		\item[] {$q_{i\omega}^S$}\qquad \quad \ {Spot market electricity quantity [MWh]}
  		\item[] {$\varepsilon_i^F$}\qquad \quad \ \ {Emission allowances for generator $i$ in the futures market [ton CO$_2$]}
  		\item[] {$\varepsilon_{i\omega}^S$}\qquad \quad \ {Emission allowances for generator $i$ in the spot market [ton CO$_2$]}
  		\item[]  {$\theta_{k\omega},\lambda_k,\mu_{k\omega}, \nu_k$ }{dual variables where $k\in I\cup J$} 
  	\end{itemize}
 
  \end{itemize}
    \subsection{The General Model (GM)}
    In the general model, we consider a set of oligopolistic conventional generators ($I$) and a set of RES generators ($J$) that compete to supply a homogeneous product (electricity) in a two-stage electricity market. In the first stage, conventional and renewable generators simultaneously choose the quantities $q_{k}^F$ where $k\in I\cup J,$ to sell in the futures market with futures price $P^F$, together with allowance futures commitment $\varepsilon_i^F$ (only for conventional generators), with futures CO$_2$ price $P^{F_{CO_2}}$. In the second stage, generators participate in a spot market where, for each scenario $\omega,$ the amount of energy $q_{k\omega}^S$, is delivered at the spot price $P_{\omega}^S$. Additionally, we assume conventional generators auction their shortage/surplus EUAs, $\varepsilon_{i\omega}^S,$ with spot CO$_2$ price, $P_{\omega}^{S_{CO_2}}.$ Note that $P_{\omega}^{S_{CO_2}}$ and $P^{F_{CO_2}}$ are exogenous parameters of the model.
    
    Considering the emissions market, the cost (income) of buying (selling) the shortage (surplus) of emission allowances is added to the total profit function of conventional generators. We assume no grandfathering allowance allocations so that generators can only auction their emission permits \citep{brown2017electricity,samadi2018analyzing,el2019investigating}. 
    
    In a two-stage stochastic programming approach where futures and spot markets are considered, the profit for conventional generator $i,$ and scenario $\omega,$ is expressed as follows:
    \begin{equation}\label{E1}
    \Pi_{i\omega}=P^Fq_i^F+P^S_{\omega}q_{i\omega}^S-C_{i\omega}(q_i^F,q_{i\omega}^S) -P^{F_{CO_2}}\varepsilon_i^F+P_{\omega}^{S_{CO_2}}(\varepsilon_{i}^F-\hat{\eta_{i\omega}}(q_i^F+q_{i\omega}^S))
    \end{equation}
    where $C_{i\omega}(q_i^F,q_{i\omega}^S)=a_{i\omega}+ b_{i\omega} (q_i^F+q_{i\omega}^S)+\frac{1}{2}c_{i\omega}(q_i^F+q_{i\omega}^S)^2$ is a quadratic production cost function\footnote{Quadratic cost functions accurately model the actual response of conventional generators where fuel is oil, coal and gas etc \citep{theerthamalai2010effective}. Renewable energy sources are without cost function because the fuel that drives its power generation is without price.} (concave and non-decreasing in input prices). $a_{i\omega}\geq 0$, $b_{i\omega}\geq 0$ and $c_{i\omega}\ge 0$ are the cost coefficients of conventional power generation $i,$ and scenario $\omega$. $P^{F_{CO_2}}$ and $P_{\omega}^{S_{CO_2}}$ are emission allowance prices in \euro/ton CO$_2$ in the futures market and in the spot market, respectively. Note that allowance price in the futures market and spot market are exogenous to the model. The total amount of CO$_2$ emissions for generator $i,$ at scenario $\omega,$ is:
	 \begin{equation}
    \varepsilon_{i\omega}^{tot}= \hat{\eta_{i\omega}}\sum_{i\in{I}}(q_{i\omega}^S+q_i^F)
    \end{equation}
        The remaining emissions allowance that can be traded in the spot market is expressed as:
    \begin{equation}\label{E2}
    \varepsilon_{i\omega}^S= \hat{\eta_{i\omega}}\sum_{i\in{I}}(q_{i\omega}^S+q_i^F)-\varepsilon_{i}^F \quad \forall i,\forall\omega
    \end{equation}
    where $\varepsilon_{i\omega}^S$ is the emissions allowance of generator $i$ in the spot market and $\hat{\eta_{i\omega}}$ is the emission intensity factor for generator $i$ and scenario $\omega$. If $\varepsilon_{i\omega}^S>0$, then generator $i$ has shortage of emissions and surplus if $\varepsilon_{i\omega}^S<0.$
 The profit for RES generator $j$, per scenario $\omega$, is expressed in (\ref{E_2}), where the production cost is assumed to be zero. 
    \begin{equation}\label{E_2}
    \Pi_{j\omega}=(P^F-P_{\omega}^S)q_j^F+P_{\omega}^SQ_{j\omega} 
    \end{equation}
    where $Q_{j\omega}=q_j^F+q_{j\omega}^S$ is a stochastic parameter representing the total production for RES generator $j$ and scenario $\omega$ which is sold both in the spot and futures markets. Indeed, the energy that can be traded in the spot market is $q_{j\omega}^S=Q_{j\omega}-q_j^F.$ As the occurrence of renewable spillage is a rare situation in most large power systems (it only occurs a few hours per year), we assume that $Q_{j\omega}$ is always dispatched at no costs or emissions with no possibility of spillage.
   
    The spot price-inverse demand curve per scenario $\omega,$ is endogenously defined as follows: 
    \begin{equation}
    P_{\omega}^S =\gamma^S_{\omega}-\beta^S_{\omega}\left(\sum_{i\in {I}} (q_{i\omega}^S+q_{i}^F)+\sum_{j\in {J}}Q_{j\omega} \right)
    \end{equation}
    with $\gamma_{\omega}^S$ and $\beta_{\omega}^S$ being positive for the price-demand function to be well behaved\footnote{The demand for electricity is assumed to exhibit the fundamental law of demand, that is, keeping other things constant, the quantity demanded decreases when the price increases.}.
    The spot market price can be simplified as follows:
    \begin{equation}\label{E4}
    P_{\omega}^{S}=\hat{\gamma}_{\omega}^{S}-\beta_{\omega}^S\sum_{i\in {I}} (q_{i\omega}^S+q_i^F)\quad \forall\omega
    \end{equation}
    where $$\hat{\gamma}_{\omega}^{S}=\gamma_{\omega}^{S}-\beta_{\omega}^S\sum_{j\in {J}} Q_{j\omega}\quad\forall\omega.$$
   Similarly, the inverse demand curve in the futures market is expressed as:
    \begin{equation}
    P^F=\gamma^F-\beta^F\left(\sum_{i\in {I}} q_{i}^F+\sum_{j\in {J}} q_{j}^F\right).
    \end{equation}
     \subsubsection{Second Stage: Spot Market Equilibrium}
    At this stage we assume that the futures market has already been settled so that we seek to characterize the spot market equilibrium, per scenario $\omega.$ Hence, the futures (first-stage) decisions variables are considered fixed at this stage. This is specified in \cref{as_1}.
    \begin{ass}\label{as_1}
        The futures market quantity $q_k^F,$ for $k\in I\cup J$ has already been committed with the futures price $P^F,$ and in the spot market all firms simultaneously maximize their profits. Moreover, each generator has an estimation (conjecture) of the impact that its production, $q_{k\omega}^S$ and $q_k^F,$ may have in the spot market price and rival quantities.
     \end{ass}
    \begin{prop}\label{E5}
        Given \cref{as_1} and the futures market quantities $q_k^F, \forall k$, at each scenario $\omega$, the equilibrium spot market price is expressed as:
        \begin{eqnarray}\label{E6}
        P_{\omega}^S&=&\varphi_{\omega}\left[\hat{\gamma}_{\omega}^{S}+\beta_{\omega}^S\sum_{i\in {I}} \tau_{i\omega}\left(b_{i\omega}+c_{i\omega}q_{i}^F+\hat{\eta_{i\omega}}P_{\omega}^{S_{CO_2}}\right)-\beta_{\omega}^S\sum_{i\in {I}}q_i^F\right]
        \end{eqnarray}
        where $\tau_{i\omega}=\frac{1}{\beta_{\omega}^S(1+\delta_i)+c_{i\omega}}$ and $\varphi_{\omega}=\frac{1}{1+\beta_{\omega}^S\sum_{i\in {I}}\tau_{i\omega}}$. The parameter $\delta_i=\sum_{k\ne i}^{I}\frac{\partial q_{k\omega}^S}{\partial q_{i\omega}^S}$ measures  the level of competition of generator $i$ in the spot market, as analyzed in \cite{lindh1992inconsistency}, i.e, $\delta_i =-1$ represents perfect competition and $\delta_i =0$ characterizes Cournot competition.
    \end{prop}
    \begin{proof}\label{thm_1}
        In the spot market profit for generator $i$ and scenario $\omega$ is expressed as:
        \begin{eqnarray}
        \Pi_{i\omega}^S(q_{i\omega}^S,q_{-i\omega}^S)=P^S_{\omega}q_{i\omega}^S-C_{i\omega}(q_i^F,q_{i\omega}^S) +P_{\omega}^{S_{CO_2}}(\varepsilon_{i}^F-\hat{\eta_{i\omega}}(q_i^F+q_{i\omega}^S)) 
        \end{eqnarray}
        The optimal quantity $q_{i\omega}^S$ is obtained by maximizing profit $\Pi_{i\omega}.$ The first order optimality condition (FOC) with respect to $q_{i\omega}^S$ are:
        \begin{eqnarray}
        \frac{\partial\Pi_{i\omega}}{\partial q_{i\omega}^S}&=&\frac{\partial P_{\omega}^S}{\partial q_{i\omega}^S}q_{i\omega}^S+P_{\omega}^S-b_{i\omega}-c_{i\omega}(q_{i}^F+q_{i\omega}^S)+P_{\omega}^{S_{CO_2}}\frac{\partial (\varepsilon_{i}^F-\hat{\eta_{i\omega}}(q_i^F+q_{i\omega}^S))}{\partial q_{i\omega}^S}\nonumber
        \end{eqnarray}
        \begin{eqnarray}
            \label{E7}
        =-\beta_{\omega}^S(1+\delta_i)q_{i\omega}^S+P_{\omega}^S-b_{i\omega}-c_{i\omega}(q_{i}^F+q_{i\omega}^S)
        -\hat{\eta_{i\omega}}P_{\omega}^{S_{CO_2}}=0
        \end{eqnarray}
        where 
        \begin{eqnarray}\label{EQ_1}
        \frac{\partial P_{\omega}^S}{\partial q_{i\omega}^S}&=&-\beta_{\omega}^S\left(1+\sum_{(k \in {I\cup J)\ne i}}\frac{\partial q_{k\omega}^S}{\partial q_{i\omega}^S} \right)=-\beta_{\omega}^S(1+\delta_i)\quad\text{and}\\
        \delta_i&=&\sum_{k\ne i}\frac{\partial q_{k\omega}^S}{\partial q_{i\omega}^S}  
        \end{eqnarray}
        Substituting (\ref{EQ_1}) into (\ref{E7}), we have:
        \begin{eqnarray}
        -\beta_{\omega}^S(1+\delta_i)q_{i\omega}^S-c_{i\omega}q_{i\omega}^S =
        - P_{\omega}^S+b_{i\omega}+c_{i\omega}q_{i}^F+\hat{\eta_{i\omega}}P_{\omega}^{S_{CO_2}}
        \end{eqnarray}
        \begin{eqnarray}\label{RR}  
        q_{i\omega}^S\left[\beta_{\omega}^S(1+\delta_i)+c_{i\omega}\right] = P_{\omega}^S-b_{i\omega}-c_{i\omega}q_{i}^F-\hat{\eta_{i\omega}}P_{\omega}^{S_{CO_2}}
        \end{eqnarray} 
        and after some rearrangements, we see that (\ref{RR}) is equivalent to:
        \begin{equation} 
        \label{E_1}
        q_{i\omega}^S =\tau_{i\omega}\left[P_{\omega}^S-b_{i\omega}-c_{i\omega}q_{i}^F-\hat{\eta_{i\omega}}P_{\omega}^{S_{CO_2}}\right]
        \end{equation}
        where $$\tau_{i\omega}= \frac{1}{\beta_{\omega}^S(1+\delta_i)+c_{i\omega}}.$$
        By substituting $q_{i\omega}^S$ of $(\ref{E_1})$ into $P_{\omega}^S$ formula in (\ref{E4}):
        \begin{eqnarray}\label{E8}
        P_{\omega}^{S}=\hat{\gamma}_{\omega}^{S}+\beta_{\omega}^S\sum_{i\in {I}} \tau_{i\omega}\left[-P_{\omega}^S+b_{i\omega}+c_{i\omega}q_{i}^F+\hat{\eta_{i\omega}}P_{\omega}^{S_{CO_2}}\right]-\beta_{\omega}^S\sum_{i\in {I}}q_i^F
        \end{eqnarray}
    and then:
        \begin{eqnarray}\label{E9}
        P_{\omega}^{S}+\beta_{\omega}^S\sum_{i\in {I}}\tau_{i\omega}P_{\omega}^S=\hat{\gamma_{i\omega}}^S+\beta_{\omega}^S\sum_{i\in {I}} \tau_{i\omega}\left[b_{i\omega}+c_{i\omega}q_{i}^F+\hat{\eta_{i\omega}}P_{\omega}^{S_{CO_2}}\right]
        -\beta_{\omega}^S\sum_{i\in {I}}q_i^F.
        \end{eqnarray}
        Thus, the equilibrium spot price is stated as follows:
        \begin{eqnarray}\label{E10}
        P_{\omega}^S=\varphi_{\omega}\left[\hat{\gamma}_{\omega}^{S}+\beta_{\omega}^S\sum_{i\in {I}} \tau_{i\omega}\left(b_{i\omega}+c_{i\omega}q_{i}^F+\hat{\eta_{i\omega}}P_{\omega}^{S_{CO_2}}\right)-\beta_{\omega}^S\sum_{i\in {I}}q_i^F\right]
        \end{eqnarray}
        where $\varphi_{\omega}=\frac{1}{1+\beta_{\omega}^S\sum_{i\in {I}}\tau_{i\omega}}.$
    \end{proof}
    \begin{prop}\label{pr_1}
        From (\ref{E_1}) and (\ref{E10}), at each scenario $\omega$ the equilibrium quantity $q_{i\omega}^S$ in the spot market is expressed as:
        \begin{eqnarray}\label{E11}
        q_{i\omega}^S &=&\tau_{i\omega}\varphi_{\omega}\left[\hat{\gamma}_{\omega}^{S}+\beta_{\omega}^S\sum_{i\in {I}}\tau_{i\omega}\left(b_{i\omega}+c_{i\omega}q_{i}^F+\hat{\eta_{i\omega}}P_{\omega}^{S_{CO_2}}\right)-\beta_{\omega}^S\sum_{i\in {I}}q_i^F\right]+\nonumber\\
        &-&\tau_{i\omega}\left[b_{i\omega}+c_{i\omega}q_{i}^F+\hat{\eta_{i\omega}}P_{\omega}^{S_{CO_2}}\right]  \quad\forall i,\forall\omega.
        \end{eqnarray}
    \end{prop}
    \begin{proof}
    We can combine (\ref{E_1}) and (\ref{E10}) to express the equilibrium quantity in terms of futures decision variables as follows: 
    \begin{equation}\label{E120} 
        q_{i\omega}^S =\tau_{i\omega}\left[P_{\omega}^S-b_{i\omega}-c_{i\omega}q_{i}^F-\hat{\eta_{i\omega}}P_{\omega}^{S_{CO_2}}\right] \quad\forall i, \forall\omega.
        \end{equation}
    Since we have the equilibrium spot market price in (\ref{E10}), substituting it into (\ref{E120}) gives:
            \begin{eqnarray}
        q_{i\omega}^S &=&\tau_{i\omega}\varphi_{\omega}\left[\hat{\gamma}_{\omega}^{S}+\beta_{\omega}^S\sum_{i\in {I}}\tau_{i\omega}\left(b_{i\omega}+c_{i\omega}q_{i}^F+\hat{\eta_{i\omega}}P_{\omega}^{S_{CO_2}}\right)-\beta_{\omega}^S\sum_{i\in {I}}q_i^F\right]+\nonumber \\
        &-&\tau_{i\omega}\left[b_{i\omega}+c_{i\omega}q_{i}^F+\hat{\eta_{i\omega}}P_{\omega}^{S_{CO_2}}\right] \quad\forall i, \forall\omega \nonumber 
        \end{eqnarray}
    \end{proof}
    \begin{prop}\label{pr_2}
Given the spot emissions in (\ref{E2}), the pre-existing allowance futures commitment, $\varepsilon_i^F,$ and the knowledge of $q_{i\omega}^S$ from \cref{pr_1}, the equilibrium amount of shortage/surplus of emission allowances for each generator at each scenario in the spot market is expressed as:
\begin{eqnarray}\label{E12}
\varepsilon_{i\omega}^S&=&\hat{\eta_{i\omega}}\sum_{i\in {I}}q_i^F+\hat{\eta_{i\omega}}\sum_{i\in {I}}\tau_{i\omega}\varphi_{\omega}\left[\hat{\gamma}_{\omega}^{S}+\beta_{\omega}^S\sum_{i\in {I}}\tau_{i\omega}\left(b_{i\omega}+c_{i\omega}q_{i}^F+\hat{\eta_{i\omega}}P_{\omega}^{S_{CO_2}}\right)\right]-\varepsilon_i^F+\nonumber\\
 &-&\hat{\eta_{i\omega}}\sum_{i\in {I}}\tau_{i\omega}\varphi_{\omega}\beta_{\omega}^S\sum_{i\in {I}}q_i^F-\hat{\eta_{i\omega}}\sum_{i\in {I}}\tau_{i\omega}\left[b_{i\omega}+c_{i\omega}q_{i}^F+\hat{\eta_{i\omega}}P_{\omega}^{S_{CO_2}}\right] \quad\forall i,\forall\omega
        \end{eqnarray}
    \end{prop}
     \begin{proof}
    The spot market emissions are expressed as $\varepsilon_{i\omega}^S=\hat{\eta_{i\omega}}\sum_{i\in {I}}(q_{i\omega}^S+q_i^F)-\varepsilon_{i}^F$ in (\ref{E2}). Expression (\ref{E12}) is obtained by substituting the spot market equilibrium quantity from \cref{pr_1} into (\ref{E2}).
    \end{proof}
Finally, as indicated above, for a given level of futures contracting, the optimal production at each scenario $\omega$ for RES generator $j$ can be expressed as  $q_{j\omega}^S=Q_{j\omega}-q_j^F.$
 \subsubsection{First Stage: Futures Market Analysis}
Now we go one step backward in time to analyze generators' futures market contracts. We can substitute the spot market equilibrium outcomes obtained in the previous section (which are parameterized in the futures decisions variables) in the profit functions expressed in (\ref{E1}) and (\ref{E_2}) for conventional and RES technologies, respectively.
    \begin{ass}\label{ass_E1}
Each generator $k$ has an estimation (conjecture) of the impact that its production $q_{k}^F$ may have in the futures price and competitors' production. 
    \end{ass}
Notice that this assumption is important to compute $\frac{\partial \Pi_{k\omega}}{\partial q_{k}^F}$ which in turn depends on $\psi_k=\frac{\partial q_{-k}^F}{\partial q_k^F},$ which is also a pre-specified parameter used to model different levels of competitions in the futures market.
Let us first derive the partial derivatives of the profit function with respect to $q_i^F,\varepsilon_i^F$, and $q_j^F$, respectively, which play a key role in the following section.
 \begin{subequations}\label{E13}
        \begin{align}
        \frac{\partial \Pi_{i\omega}}{\partial q_i^F}&=\frac{\partial P_{\omega}^S}{\partial q_i^F}q_{i\omega}^S+P_{\omega}^S\frac{\partial q_{i\omega}^S}{\partial q_i^F}
        -b_{i\omega}\left(1+\frac{\partial q_{i\omega}^S}{\partial q_i^F} \right)-c_{i\omega}\left[ q_i^F+q_{i\omega}^S \right] \left(1+\frac{\partial q_{i\omega}^S}{\partial q_i^F} \right) \nonumber\\
        &+\frac{\partial P^F}{\partial q_i^F}q_i^F-\hat{\eta_{i\omega}}P_{\omega}^{S_{CO_2}}\left(1+\frac{\partial q_{i\omega}^S}{\partial q_i^F}\right)+P^F  \quad\forall i, \forall\omega\\ 
        \frac{\partial \Pi_{i\omega}}{\partial \varepsilon_i^F}&=P_{\omega}^{S_{CO_2}}-P^{F_{CO_2}}  \quad\forall i, \forall\omega\\
        \frac{\partial \Pi_{j\omega}}{\partial q_j^F}&=\left(\frac{\partial P^F}{\partial q_j^F}-\frac{\partial P_{\omega}^S}{\partial q_j^F} \right)q_j^F-P_{\omega}^S+\frac{\partial P_{\omega}^S}{\partial q_j^F}Q_{j\omega} +P^F  \quad  \forall j, \forall\omega
        \end{align}
    \end{subequations}    
    Moreover, considering the spot market equilibrium outcomes derived in the aforementioned section, we can also compute the following partial derivatives:
    \begin{subequations}\label{EE_1}
        \begin{align}
        \frac{\partial P^F}{\partial q_i^F}&=-\beta^F\left( 1+\sum_{(k \in {I\cup J)\ne i}}\frac{\partial q_{k}^F}{\partial q_i^F}+\sum_{j\in {J}}\frac{\partial q_{j}^F}{\partial q_i^F}\right)=-\beta^F(1+(I+J-1)\psi_i)  \quad\forall i\\
        \frac{\partial P_{\omega}^S}{\partial q_i^F}& =\varphi_{\omega}\beta_{\omega}^S\left[  -(1+(I-1)\psi_i)  + \varphi_{\omega}c_{i\omega}\tau_{i\omega}+\sum_{k\ne i} c_{k\omega}\psi_k\tau_{k\omega}\right]  \quad\forall i,\forall\omega\\
        \frac{\partial q_{i\omega}^S}{\partial q_i^F} &=\tau_{i\omega}\left(\frac{\partial P_{\omega}^S}{\partial q_{i}^F}-c_{i\omega}\right) \quad\forall i, \forall\omega\\
        \frac{\partial P^F}{\partial q_j^F}&=-\beta^F(1+(I+J-1)\psi_j) \quad \forall j\\
        \frac{\partial P_{\omega}^S}{\partial q_j^F} &=0 \quad \forall j,\forall\omega\\ \\
        \frac{\partial q_{\omega}^S}{\partial q_j^F} &=-1 \quad \forall j,\forall\omega.\\ 
        \end{align}
    \end{subequations}
    \subsubsection{Risk Averse Futures Market Equilibrium}\label{Sec_2}
As indicated, by replacing the equilibrium spot price ($P^S_{\omega}$) and quantities (${q^S_{i\omega}}$ and ${q^S_{j\omega}}$) in the profit functions $\Pi_{i\omega}$ and $\Pi_{j\omega}$, we can parameterize profits in the futures decision variables:
    \begin{subequations}\label{R_1}
        \begin{align}
        &\Pi_{i\omega}=\Pi_{i\omega}(q^F_i,q^F_{-i},\varepsilon_i^F,\varepsilon_{-i}^F, q^F_{j\in J}) \quad\forall i, \forall\omega\\
        &\Pi_{j\omega}=\Pi_{j\omega}(q^F_j,q^F_{-j},q^F_{i\in I})  \quad \forall j, \forall\omega
        \end{align}
    \end{subequations}
The profit maximization problem solved by the risk averse generators is formulated in \eqref{E14} where $\sigma_{k\omega}$ is the probability assigned by generator $k\in I\cup J$ to scenario $\omega$. $1-\alpha$ represents the level of significance associated with the CVaR (linear formulation introduced by \cite{rockafellar2010buffered}). 
    Therefore, risk averse problem solved by each generator is:
    \begin{subequations}\label{E14}
        \begin{align}
        \displaystyle \max_{\xi_k, \eta_{k\omega}, q_k^F,\varepsilon_i^F} & {(1-\phi)\sum_{\omega\in {\Omega}}\sigma_{k\omega}\Pi_{k\omega}}+\phi \left[\xi_k-\frac{1}{ 1-\alpha}\sum_{\omega\in {\Omega}}\sigma_{k\omega}\eta_{k\omega} \right]\\
        &\textrm{s.t.}\nonumber\\
        &\eta_{k\omega}+\Pi_{k\omega}-\xi_k\geq 0  {:\mu_{k\omega}} \quad \forall k, \forall\omega\\
        & \eta_{k\omega}\geq 0 {:\theta_{k\omega}} \quad \forall k, \forall\omega\\
        & q_{k}^{F_{min}}\leq q_{k}^{F}\leq q_{k}^{F_{max}} {:\nu_{k}^{min},\nu_{k}^{max}}\quad \forall k\label{D_2}\\
        & \varepsilon_k^{F_{min}}\leq \varepsilon_k^{F}\leq \varepsilon_k^{F_{max}} {:\lambda_{k}^{min},\lambda_{k}^{max}}\quad \forall k\label{d_3}
        \end{align}
    \end{subequations}
     The objective function is expressed as the expected profit (first term) and the CVaR (second term) multiplied by a risk aversion parameter $\phi\in [0,1]$. $\phi$ regulates the balance between expected profits and the CVaR for a given confidence level $\alpha$. In other words, $\phi=0$ represents maximization of the expected profits (risk neutral agent), and $\phi=1$ corresponds to the maximization of the most risk averse setting in which all the weight in the objective function is placed on the CVaR. The optimal value of the auxiliary variable $\xi_{k}$ corresponds to the VaR and $\eta_{k\omega}$ represents the positive discrepancy between VaR and each profit scenario. Constraints (\ref{D_2}) and (\ref{d_3}) enforce the lower and upper bounds for generation and emission, respectively, of generator $k$.
    $\theta_{k\omega},\nu_{k}^{min},\nu_{k}^{max},\lambda_{k}^{min},$ and $\lambda_{k}^{max}$ are the Lagrangian ($\mathcal{L}$) multipliers associated with the constraints of CVaR, minimum and maximum quantities of generation and emissions which are treated as variables in the equilibrium model. Note that $\Pi_{k\omega}$ has a quadratic expression (\ref{R_1}) so that problem (\ref{E14}) is indeed a quadratic optimization problem with quadratic constraints.
    
    Finally, the market equilibrium model is obtained by solving simultaneously (\ref{E14}) for all generators $k\in I\cup J$ after replacing (\ref{E14}) by its associated KKT system of optimality conditions. In particular, the KKT conditions associated with generator $k$ is:
    \begin{subequations}\label{eq_5}
        \begin{align}
        &\frac{\partial\mathcal{L}}{\partial q_k^F}= {-(1-\phi)\sum_{\omega\in {\Omega}}\sigma_{k\omega}\frac{\partial\Pi_{k\omega}}{\partial q_k^F}}-\sum_{\omega\in {\Omega}}\mu_{k\omega}\frac{\partial\Pi_{k\omega}}{\partial q_k^F}-\nu_k^{min}+\nu_k^{max}=0 \quad \forall\omega \label{second_a}\\
        &\frac{\partial\mathcal{L}}{\partial \varepsilon_k^F}= {-(1-\phi)\sum_{\omega\in {\Omega}}\sigma_{k\omega}\frac{\partial\Pi_{k\omega}}{\partial \varepsilon_k^F}}-\sum_{\omega\in {\Omega}}\mu_{k\omega}\frac{\partial\Pi_{k\omega}}{\partial \varepsilon_k^F}-\lambda_k^{min}+\lambda_k^{max}=0 \quad\forall\omega \label{second_b}\\
        &\frac{\partial\mathcal{L}}{\partial \eta_{k\omega}}=\phi\frac{1}{1-\alpha}\sigma_{k\omega}-\mu_{k\omega}-\theta_{k\omega}=0 \quad \forall\omega\label{second_c}\\
        & \frac{\partial\mathcal{L}}{\partial \xi_{k}}=-\phi + \sum_{\omega\in {\Omega}}\mu_{k\omega}=0\quad \forall\omega\label{second_d}\\
        &0\leq \eta_{k\omega}+\Pi_{k\omega}-\xi_k \perp \mu_{k\omega}\geq 0 \quad  \forall\omega\label{second_e}\\
        &0\leq \eta_{k\omega} \perp \theta_{k\omega}\geq 0 \quad \forall\omega\label{second_f}\\
        &0\leq q_{k}^{F}-q_{k}^{F_{min}} \perp \nu_{k}^{min}\geq 0  \label{second_g}\\
        &0\leq q_{k}^{F_{max}}-q_{k}^{F} \perp \nu_{k}^{max}\geq 0 \label{second_h}\\
        &0\leq \varepsilon_{k}^{F}-\varepsilon_{k}^{F_{min}} \perp \lambda_{k}^{min}\geq 0 \label{second_i}\\
        &0\leq \varepsilon_{k}^{F_{max}}-\varepsilon_{k}^{F} \perp \lambda_{k}^{max}\geq 0. \label{second_j}
        \end{align}
    \end{subequations}
    For computational purposes, it is convenient to reformulate the system of equations in (\ref{eq_5}) $\forall k$ as nonlinear optimization problem that minimizes sum of  complementarity constraints (of a form $xy=0$) in (\ref{eq_5}) subject to the remaining set of inequality and equality constraints in (\ref{eq_5}).
    \begin{subequations}\label{E_15}
    \begin{align}
    & \text{Min} \sum_{i\in {I}}\sum_{\omega\in {\Omega}}\mu_{i\omega}(\eta_{i\omega}+\Pi_{i\omega}-\xi_{i})+\sum_{j\in {J}}\sum_{\omega\in {\Omega}}\mu_{j\omega}(\eta_{j\omega}+
    \Pi_{j\omega}-\xi_{j})+\sum_{i\in {I}}\sum_{\omega\in {\Omega}}\eta_{i\omega}\theta_{i\omega}\nonumber\\
    &+\sum_{j\in {J}}\sum_{\omega\in {\Omega}}\eta_{j\omega}\theta_{j\omega}
    +\sum_{i\in {I}}\left(q_i^F-q_{i}^{F_{min}}\right)\nu_{i}^{min} 
    +\sum_{j\in {J}}\left(q_j^F-q_{j}^{F_{min}} \right)\nu_{j}^{min}
    \nonumber\\
    &+\sum_{i\in {I}}\left(q_{i}^{F_{max}}-q_i^F \right)\nu_{i}^{max}+\sum_{j\in {J}}\left(q_{j}^{F_{max}}-q_j^F\right)\nu_{j}^{max}+\sum_{i\in {I}}\left(\varepsilon_i^F-\varepsilon_{i}^{F_{min}}\right)\lambda_{i}^{min}\nonumber\\
    &+\sum_{i\in {I}}\left(\varepsilon_{i}^{F_{max}}-\varepsilon_i^F\right)\lambda_{i}^{max}\\
    &\text{subject to}\nonumber\\
    & \text{equalities}\quad  (\ref{second_a})-(\ref{second_d})\\ &\text{inequalities}\quad  (\ref{second_e})-(\ref{second_j})\\
    &(\ref{E1}),(\ref{E2}),(\ref{E6}),(\ref{E11}),(\ref{EE_1})
     \end{align}
    \end{subequations}
   where (\ref{E1}) and (\ref{E2}) are profit definitions for conventional and RES generators, respectively, whereas, (\ref{E6}) and (\ref{E11}) are equilibrium spot price and spot quantity, and (\ref{EE_1}) the partial derivatives completing the futures market.
 
    Problem (\ref{E_15}) can be tackled by using standard NLP solvers. Moreover, if a solution to (\ref{E_15}) renders an objective function value equal to $0$, then this is also the solution of system (\ref{eq_5}) \citep{leyffer2010solving}.
   
    \subsection{Spot Market with no Futures}\label{sonly}
    Finally, to complete the analysis and for comparison purposes, we derive the equilibrium market outcomes of a market with no futures trading (neither electricity nor allowances). The profit for conventional generator $i$ with no futures trading, subject to uncertain emissions auctioning, is expressed as:
    \begin{equation}
    \Pi_{i\omega}(q_{i\omega}^S,q_{-i\omega}^S) = P^S_{\omega}q_{i\omega}^S-a_{i\omega}- b_{i\omega} q_{i\omega}^S-\frac{1}{2}c_{i\omega}(q_{i\omega}^S)^2- \hat{\eta_{i\omega}}P_{\omega}^{S_{CO_2}}q_{i\omega}^S\quad \forall\omega
    \end{equation}
    and the profit for renewable generator $j$ is expressed as: 
    \begin{equation}
    \Pi_{j\omega}=P_{\omega}^SQ_{j\omega} \quad \forall\omega
    \end{equation}
    In this case, the inverse demand function is expressed as:
    \begin{eqnarray}
    P_{\omega}^S =\gamma^S_{\omega}-\beta^S_{\omega}\left(\sum_{i\in {I}} q_{i\omega}^S+\sum_{j\in {J}}Q_{j\omega}\right)\quad \forall\omega
    \end{eqnarray}
    Which can be simplified as:
    \begin{eqnarray}\label{only_1}
    P_{\omega}^{S}=\hat{\gamma}_{\omega}^{S}-\beta_{\omega}^S\sum_{i\in {I}} q_{i\omega}^S\quad \forall\omega\end{eqnarray} 
    where $$\hat{\gamma}_{\omega}^{S}=\gamma_{\omega}^{S}-\beta_{\omega}^S\sum_{j\in {J}} Q_{j\omega}  \quad \forall\omega.$$
    The equilibrium in the spot market, per scenario $\omega$, is reached when all generators $i$ maximize their profits simultaneously:
    \begin{equation}\label{S1}
    \frac{\partial\Pi_{i\omega}}{\partial q_{i\omega}^S}=\frac{\partial P_{\omega}^S}{\partial q_{i\omega}^S}q_{i\omega}^S+P_{\omega}^S-b_{i\omega}-c_{i\omega}q_{i\omega}^S-\hat{\eta_{i\omega}}P_{\omega}^{S_{CO_2}}=0 
    \end{equation}    
    Simplifying \eqref{S1} gives: 
    \begin{equation}\label{S2}
    q_{i\omega}^S=\tau_{i\omega}(P_{\omega}^S-b_{i\omega}-\hat{\eta_{i\omega}}P_{\omega}^{S_{CO_2}}) \quad\forall i,\forall\omega
    \end{equation}
    By substituting (\ref{only_1}) into (\ref{S2}), we obtain:
    \begin{equation*}\label{S3}
    P_{\omega}^S+\beta^S_{\omega}\sum_{i\in {I}}\tau_{i\omega}P_{\omega}^S=\hat{\gamma}_{\omega}^{S}+\beta^S_{\omega}\sum_{i\in {I}}\tau_{i\omega}(b_{i\omega}+\hat{\eta_{i\omega}}P_{\omega}^{S_{CO_2}}) 
    \end{equation*}
    \begin{equation}\label{S4}
    P_{\omega}^S\left(1+\beta^S_{\omega}\sum_{i\in {I}}\tau_{i\omega}\right) = \hat{\gamma}_{\omega}^{S}+\beta^S_{\omega}\sum_{i\in {I}}\tau_{i\omega} (b_{i\omega}+\hat{\eta_{i\omega}}P_{\omega}^{S_{CO_2}})
    \end{equation}
    Finally, the equilibrium clearing price per scenario $\omega$ is: 
    \begin{equation}
    \label{S5}
    P_{\omega}^S=\varphi_{\omega}\left[\hat{\gamma}_{\omega}^{S}+\beta_{\omega}^S\sum_{i\in {I}}\tau_{i\omega} (b_{i\omega}+\hat{\eta_{i\omega}}P_{\omega}^{S_{CO_2}}) \right]
    \end{equation}
    By substituting \eqref{S5} into \eqref{S2}, we can get equilibrium quantity for conventional generators as:
    \begin{equation}\label{S6}
    q_{i\omega}^S=\tau_{i\omega}\left[\varphi_{\omega}\left(\hat{\gamma}_{\omega}^{S}+\beta_{\omega}^S\sum_{i\in {I}}\tau_{i\omega} (b_{i\omega}+\hat{\eta_{i\omega}}P_{\omega}^{S_{CO_2}}) \right)-b_{i\omega}-\hat{\eta_{i\omega}}P_{\omega}^{S_{CO_2}}\right]\quad\forall i,\forall\omega
    \end{equation}
    \begin{equation}\label{S7}
    \varepsilon_{i\omega}^S=\hat{\eta_{i\omega}}\sum_{i\in {I}}q_{i\omega}^S \quad\forall i,\forall\omega
    \end{equation}
    Finally, the emission in the spot market for generator $i$, per scenario $\omega$, is expresses as:
    \begin{equation}\label{S8}
    \varepsilon_{i\omega}^S=\hat{\eta_{i\omega}}\sum_{i\in {I}}\tau_{i\omega}\left[\varphi_{\omega}\left(\hat{\gamma}_{\omega}^{S}+\beta_{\omega}^S\sum_{i\in {I}}\tau_{i\omega} (b_{i\omega}+\hat{\eta_{i\omega}}P_{\omega}^{S_{CO_2}}) \right)-b_{i\omega}-\hat{\eta_{i\omega}}P_{\omega}^{S_{CO_2}}\right].
    \end{equation}
    	\section{Numerical Analysis}\label{S_4} 
	We have analytically derived the game-theoretical equilibrium models (the general model and the spot only market model). It should be noted that modeling emissions auctioning in oligopolistic electricity market with financial derivatives is the key focus of this paper. In order to evaluate the efficiency of these models, we carried out various numerical studies.
	\subsection{Data}
This section provides numerical examples to show the performance of the proposed market equilibrium models. We calibrate the data to get working parameters which approximate the behavior of a realistic electricity market with contracts and emissions auctioning. We compare the parameters with realistic data from the Spanish electricity market \citep{OMIE2020} that combines futures contracts and pool market structures. We also use the EEX\footnote{EEX is a leading exchange platforms in the European power market that offers trading in power derivatives for \euro-denominated cash-settled futures contracts for 20 European power markets across Europe.} \citep{EEX2020} to compare the EUA futures contracting prices.
Electricity generators in the 31 countries under EU ETS can sign futures contracts (quantity, price, and delivery date) for electricity and emission allowances. 

Our model considers three conventional generators and one large RES generator participating in the electricity and emission trading market. To supply the aggregate demand, we are considering large generators whose total output is the sum of several units with different technologies. Hence, based on the Central Limit Theorem we can reasonably assume that the random parameters (renewable output, generation costs and demand curve) can be modeled by normal distribution. With that, we examine the impact of renewable penetration and CO$_2$ price variation on prices (futures and expected spot prices), quantities (futures and spot), and profits (for conventional and RES). 
    
We randomly generate the quadratic cost parameters (which is consistent with the approach in \cite{goudarzi2017smart,ruiz2012equilibria,oliveira2013contract}). For the emission intensity, our parameters are consistent with \cite{farhat2010greenhouse}. The mean values and coefficient (CV) of variations of cost parameters, emission intensity factors and the maximum generation quantities and maximum emissions auctioned for each conventional generator $i$ are presented in \cref{table_1}. Note that the standard deviation is calculated with $\sigma= \mu\times CV.$
\color{black}
\begin{table}
\centering
\begin{tabular}{lllll}
\hline
& \multicolumn{4}{l}{Generators} \\ \cline{2-5}
Parameter & $i=1$    & $i=2$    & $i=3$    & $CV$  \\ \hline
${a_{i\omega}}$& 35.00  & 45.00  & 50.00  & 0.10     \\
${b_{i\omega}}$ & 27.00  & 35.00  & 43.00  & 0.13     \\
${c_{i\omega}}$ & 0.015  & 0.008  & 0.013  & 0.15     \\
${\hat{\eta_{i\omega}}}$& 0.67   & 0.50   & 0.49   & 0.05     \\ \hline
 $\varepsilon_{i}^{F_{max}}$& 20,000 & 23,000 & 19,000 &          \\
 $q_{k}^{F_{max}}$  & 21,000 & 21,000 & 25,000 &          \\ \hline
\end{tabular}%
\caption{Mean values and coefficient of variations used to calibrate the data applied in our simulations for conventional generators. The upper limit for emission and quantity are defined.}
\label{table_1}
\end{table}

The cost parameter $a_{i\omega}$ is the cost intercept and load cost for generator $i$, and  scenario $\omega$, generated randomly with mean values equal to [35,45,50][\euro/MWh] and CV of 10\%. Similarly, $b_{i\omega}$ and $c_{i\omega}$ are generated randomly for each scenario. That means, each realization of $b_{i\omega}$ is generated from a multivariate normal distribution with mean $[27, 35, 43]$ [\euro/MWh] and a CV of 13\%, and $c_{i\omega}$ is randomly generated from a normal distribution  with mean $[0.015, 0.008, 0.013]$ [\euro/MWh$^2$] and CV of $0.015.$ The intensity factor for generator $i$, $\hat{\eta_{i\omega}}$ is randomly generated from a normal distribution with mean $[0.67,0.50,0.492]$ [CO$_2$/MWh] and CV of $0.05$.
 
The demand curve parameters for electricity market are generated by approximating the aggregated step-wise demand curve in the spot market, using the futures market intercept $\mu_{\gamma^F}=180$ and CV of $ 0.15.$ The slope is $\mu_{\beta^F}=0.005$ with CV of $0.057.$ The expected value of the parameters of the inverse spot market demand curve equal those of the futures demand curve, i.e., $\gamma^F=E[\gamma_{\omega}^S]$ and $\beta^F=E[\beta_{\omega}^S]$. From this, $\gamma_{\omega}^S$ for each scenario is randomly generated with a normal distribution as $\gamma_{\omega}^S\sim N(\gamma^F,\sigma_{\gamma^F})$ and $\beta_{\omega}^S\sim N(\beta^F,\sigma_{\beta^F})$, where CV for ${\gamma^F}$ and ${\beta^F}$ are equal to $0.15.$ and $0.005,$ respectively. For simplicity, no correlation among cost, or demand parameters are considered.
    
For the CO$_2$ price in the spot market, we take the mean futures CO$_2$ price to be  25\euro/MWh with a CV of $0.16$ and randomly generated with $\omega$ scenario.
The parameters for the futures CO$_2$ prices are chosen based on the EUA futures prices presented by EEX futures market. The data covers daily trading prices of 215 working days from January 7, 2019, to December 16, 2019 \citep{EEX2020}. Thus, for the CO$_2$ price analysis, we use the CO$_2$ price vector in the range of $[0,50]$ which corresponds with EEX EUA futures price. 

The average renewable energy total production for generator $j$ is in the range $[0,10000]$ MWh with a CV of $0.057$.

These data are generated to test the models and study the behavior of the equilibrium market outcomes, with increasing penetration of RES capacity and CO$_2$ prices. Each market equilibrium is first characterized by the optimal values of the first-stage decision variables ($P^F$ and $q_k^F$), and by the second-stage variables ($P_{\omega}^S$ and $q_{k\omega}^S$), that determine the spot equilibrium market outcomes with scenario $\omega$. The models have been implemented in JuMP version 0.21.1 \citep{ dunning2017jump} under the open source Julia programming Language version 1.5.2 \citep{bezanson2017julia}. We use Artelys Knitro solver version 12.2 \citep{byrd2006k} on an CPU E5-1650v2@3.50GHz and 64.00 GB of RAM running workstation. We analyze multiple cases with different number of equiprobable scenarios which are selected based on the overall computational performance. For instance, the risk aversion simulations using RES as a parameter for the Cournot model with  $|\Omega| = 320$ scenario is solved in 675.094 seconds of elapsed CPU time (120.38 M allocations: 4.165 GiB, 0.24\% gc time).
 \subsection{Result Presentation and Discussion}
    Based on the risk aversion problems specified in (\ref{Sec_2}), two risk profiles are demonstrated and discussed. The parameter $\phi=0$ is used for the risk neutral generator, whereas $\phi=1$ is applied for the risk averse generators (by changing the parameter $\phi$ between zero to one, the tolerable risk level of generators is adjusted). The threshold of the value at risk $\alpha$ is fixed at $0.90$.
    
As we have multiple cases, we simulate them differently and the results are reported based on different working scenarios\footnote{Most stochastic models use a limited number of scenarios for computational reasons. However, to investigate the effect of risk aversion using CVaR, this is not enough. In the real world, we need to consider low-probability/high-consequence scenarios that can have a significant effect on expected profits or welfare if they occur. Moreover, the smallest meaningful CVaR threshold level of $\alpha$ in a model with $\Omega$ scenarios is $1/\Omega$, while typical threshold values are closer to 5\% \citep{munoz2017does}. Hence, we simulate as large as 320 scenarios for the risk aversion (models with CVaR) to characterize the model realistically.}. \cref{tab_12} presents the number of working scenarios for all the cases into consideration.
\begin{table}
\centering
\begin{tabular}{|l|l|l|l|l|l|l|l|l|}
\hline
\multirow{3}{*}{$|\Omega|$} & \multicolumn{4}{c|}{Risk Neutral Model} & \multicolumn{4}{c|}{Risk Averse Model} \\ \cline{2-9} 
 &
  \multicolumn{2}{c|}{\begin{tabular}[c]{@{}c@{}}Cournot\\ Model\end{tabular}} &
  \multicolumn{2}{c|}{\begin{tabular}[c]{@{}c@{}}Competitive \\ Model\end{tabular}} &
  \multicolumn{2}{c|}{\begin{tabular}[c]{@{}c@{}}Cournot\\ Model\end{tabular}} &
  \multicolumn{2}{c|}{\begin{tabular}[c]{@{}c@{}}Competitive \\ Model\end{tabular}} \\ \cline{2-9} 
& RES    & CO$_2$    & RES    & CO$_2$    & RES    & CO$_2$    & RES    & CO$_2$   \\ \hline
100  &  & \checkmark&  & &  &  & &   \\ \hline
125 & \checkmark & & \checkmark  & & & & &  \\ \hline
200 &  &  & & \checkmark&  &&  & \\ \hline
320 &  &&  && \checkmark & \checkmark & \checkmark & \checkmark  \\ \hline
\end{tabular}%

\caption{Scenarios for each case.}
\label{tab_12}
\end{table}
\begin{table}[H]
\centering
\begin{tabular}{|l|l|l|l|l|}
\hline
$\Omega$ &  EX.$P^F$ & EX.$P_{\omega}^S$ & Con. Profit (Expected) & RES. Profit (Expected)\\ \hline
150      & 97.90 & 86.29          & 70,546      & 432,652     \\ \hline
200      & 96.28 & 86.06          & 71,056      & 432,851     \\ \hline
320      & 98.08 & 87.01          & 73,291      & 434,447     \\ \hline
350      & 98.53 & 86.65          & 72,256      & 433,660     \\ \hline
400      & 98.23 & 86.22          & 70,003      & 431,351     \\ \hline
500      & 97.67 & 85.80          & 69,254      & 428,312     \\ \hline
\end{tabular}
\caption{The simulation is run for Cournot competition with RES parameter for different number of scenarios. The sensitivity results (market outcomes) reported are the mean values for the RES parameters, from 0 to 10,000 in 1000 ranges (11 parameters). As the results show, there is no significant difference in outcomes due to the increase/decrease in the number of scenarios. This attests the number of scenarios used in our reporting works properly and change neither the concluding results nor the objective of the models.}
\label{tab_R2}
\end{table}

For the CVaR simulations, our model works even for a larger number of scenarios (for instance, 500 and more). However, if we further increase the number of scenarios for some parameter values, we will have non-zero objective function results, which show the solution is not optimal. To that end, we report the case studies with $\omega=320$ (well above the threshold ($\ge30$) for CVaR at $\alpha=0.90$), where the number of scenarios is sufficiently large to ensure the stability of the CVaR while keeping the problem tractable.

    For the sake of convenience, we present our simulation results in two parts. Since we intend to characterize the risk neutral and risk averse generators by CVaR formulation, we present the result as: 
    \begin{enumerate}
        \item \textit{Risk Neutral Generators:} with the Cournot market and the perfect competitive market structure, we discuss the general model and the spot only market. The focus is on the equilibrium prices, profits and quantities (total trading) that are presented and discussed with respect to RES penetration and CO$_2$ price parameters.
        \item \textit{Risk Averse Generators:} we use a similar approach than the risk neutral case, except, we impose the CVaR parameter to be $\phi =1$.
    \end{enumerate}
\subsection{Risk Neutral Generators Numerical Results}\label{riskneural}
This section discusses the equilibrium market for risk neutral generators using RES penetration and CO$_2$ price as parameters of interest.
\subsubsection{Sensitivity on Renewable Penetration}
\begin{figure}
        \centering
        \includegraphics[width=\textwidth]{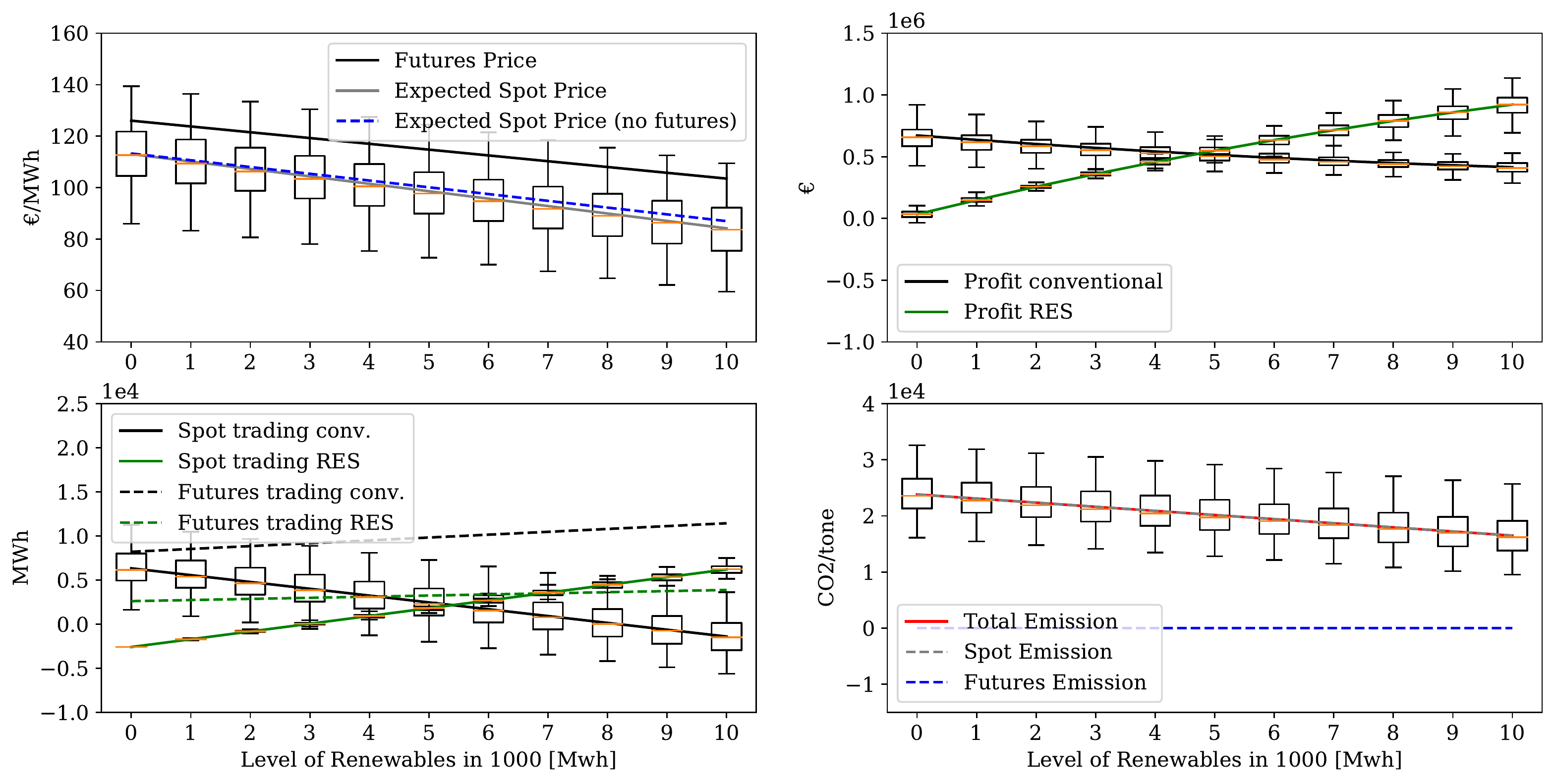}
        \caption{Risk neutral Cournot model simulation with results sensitivity analysis on RES penetration.}
        \label{fig_s1}
    \end{figure}

For the RES penetration sensitivity analysis, CO$_2$ price is fixed at its mean value of 25\euro/MWh and a CV of 0.16. 
 Let us first examine the Cournot competition equilibrium results from \cref{fig_s1}. The futures price of electricity is negatively correlated to the RES penetration (diminishing from around 126 to 103\euro/MWh). Similar behavior is observed for the expected spot price (with a range of variation of approximately $[84,113]$ \euro/MWh). This can be explained by the substitution of conventional generations for RES ones, which exhibit null production costs and curb the level of electricity prices in both markets. Expected spot electricity prices (with the presence of futures market and without the presence of futures) are overlapped for RES less than 5,000MWh. For RES $>5,000$MWh, the expected spot price in the presence of the futures market decreases faster than the expected spot price in the GM. The futures price is higher than the expected spot one.

\begin{figure}
        \centering
        \includegraphics[width=\textwidth]{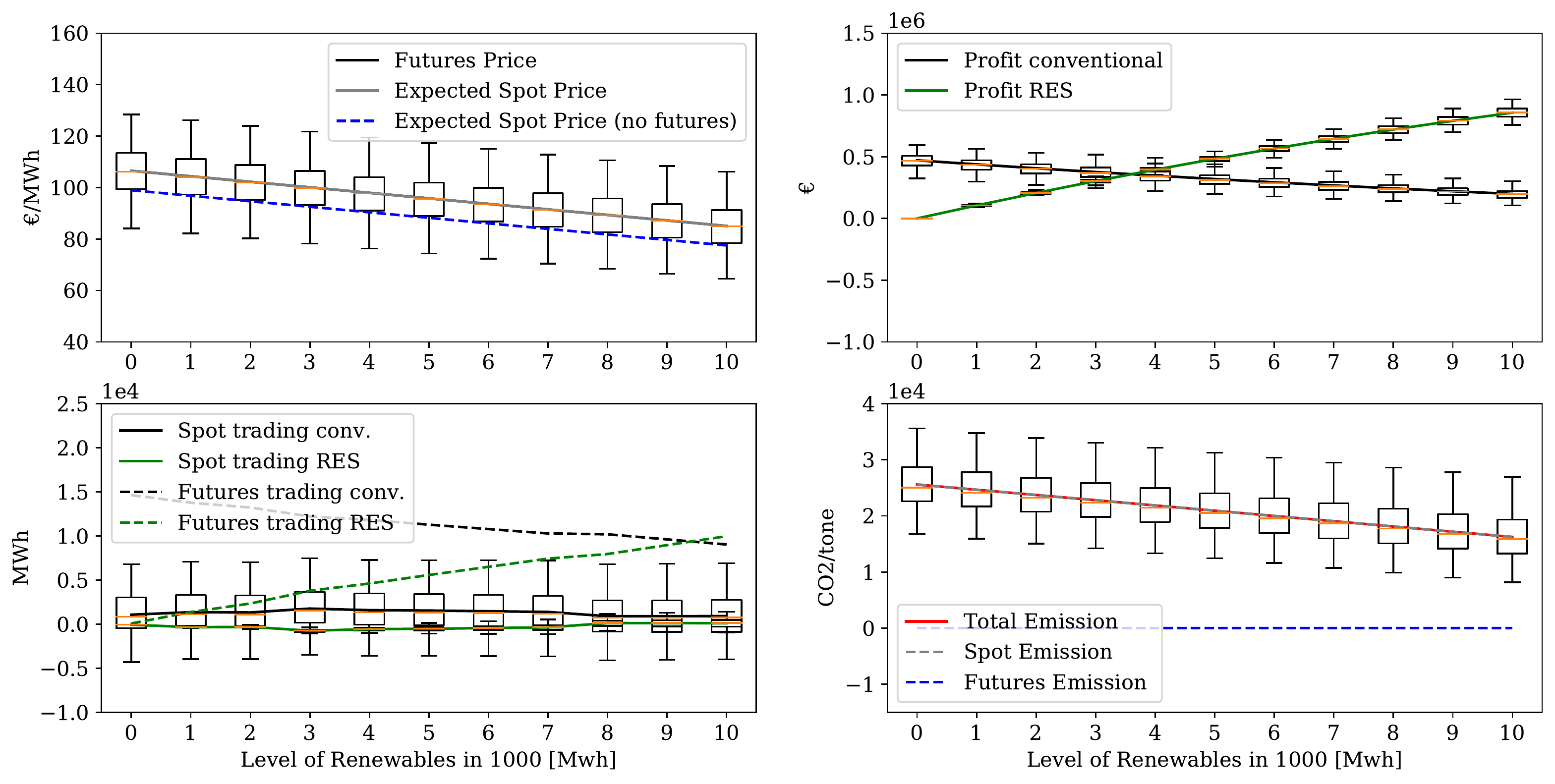}
        \caption{Risk neutral perfect competitive model with sensitivity analysis on RES penetration}
        \label{fig_s2}
    \end{figure}

 The overall RES generators' profit increases with an increasing rate with respect to RES penetration. However, conventional generators' profit decreases with RES penetration. Despite conventional generators spot market trading consistently decreases with RES penetration, their trading quantities in the futures market increases, which in turn helps them to balance their profit. Since RES penetration increases the total quantities traded by RES generators, their profit is always increasing. Conventional generators trade emissions only in the spot market. Therefore, futures market emission is almost zero with RES penetration. Note that total-vis-à-vis spot CO$_2$ emissions consistently decreases with respect to RES penetration.
With the perfect competition, we observe different behaviors.
Futures and expected spot prices without the presence of futures market overlapped and vary from 85 to 106\euro/MWh (refer \cref{tab_14} and \cref{fig_s2}). Since RES penetration decreases electricity prices, there is no incentive for conventional generators to compete in the futures market where their futures market trading decreases. This is the opposite of the Cournot competition. Comparing the two competitions, spot trading consistently decreases in the Cournot competition and slightly increases/unchanges in the perfect competition corresponding to the RES penetration. Conventional generators earn higher profit with Cournot competition than with perfect competition where futures price ranges from [103,126] to, [85,106]\euro/MWh, respectively. In the perfect competition, the reduced conventional quantities are substituted by the increase in RES generators quantities. The overall impact of RES penetration for conventional generators is higher in the Cournot competition than in the perfect competition.
\subsubsection{Sensitivity on Carbon Price Increase}
With the sensitivity analysis on CO$_2$ price, we try to examine if a EUA allowance auctioning, or charging a higher price of CO$_2$ can contribute to decarbonizing the electricity system. Therefore, we try to explore its effects on prices, profits, and quantities for  RES and conventional generators.
While simulating the model by varying the CO$_2$ price, the RES parameter is randomly generated/fixed with a mean of 5,000MWh and a CV of 0.057. To examine the results, let us start from the Cournot competition. 

The increase in CO$_2$ price has an increasing impact on the overall electricity price. Futures price increases within the range [108,120]\euro/MWh, while the expected spot price varies within the range [91,105]\euro/MWh. The expected spot market electricity price increases faster with the presence of futures market than it does with no futures market. In other words, there is no significant variation regarding the spot only electricity price (refer \cref{tab_14} for comparison). The insight is that increase in electricity price is partially to cover the increase in the marginal cost of conventional generators.

Despite the large gap between the minimum and the maximum profits based on the whiskers at CO$_2$ price 30 and 40\euro/MWh), the overall impact of CO$_2$ price increase on market outcomes is limited.

 The amount of CO$_2$ emitted by conventional generators generally decreases as their production decreases.\footnote{Note that due to the risk neutrality nature of generators with the CO$_2$ price, emissions trading in the futures market is highly volatile. Thus, to manage the inherent volatility, we plot the total emissions than the futures and spot emissions in the other cases.} The generation of sustainable energy sources contributes to reducing CO$_2$ emissions, which corroborates with total emissions reduction targets by EU ETS.

Looking at the perfect competition model, the spot and futures market prices are overlapped within the interval [85,101]\euro/MWh.  Prices are lower in the perfect competition than in the Cournot competition (for instance $[85,101]<[108,120]$ for futures price in \cref{tab_14}). This explains why conventional generators earn a lower profit than the Cournot competition, albeit trading a larger quantity in perfect competition. Part of the increase of CO$_2$ price passed into the expected spot and futures prices so that conventional generators still maintain a constant profit with the CO$_2$ price increase. Similar to the Cournot competition, in the perfect competition the profit whiskers are large.

 The negative trading quantities by conventional generators in the spot market are short selling as they trade more electricity in the futures market. Total emissions trading decreases with respect to CO$_2$ price.
    \begin{figure}
        \centering
        \includegraphics[width=\textwidth]{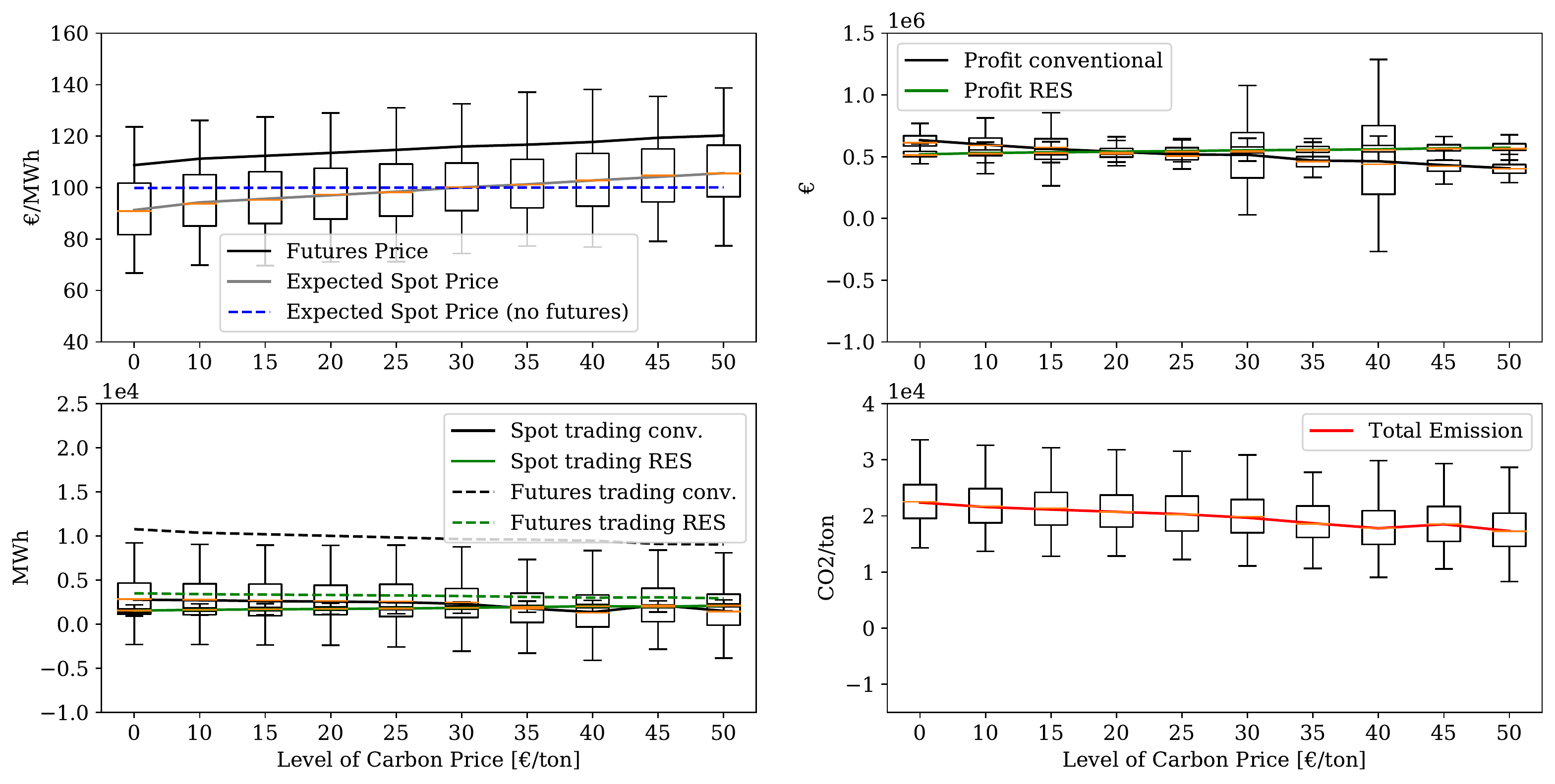}
        \caption{Risk neutral Cournot model with sensitivity on CO$_2$ price}
        \label{fig_s3}
    \end{figure}
        \begin{figure}
        \centering
        \includegraphics[width=\textwidth]{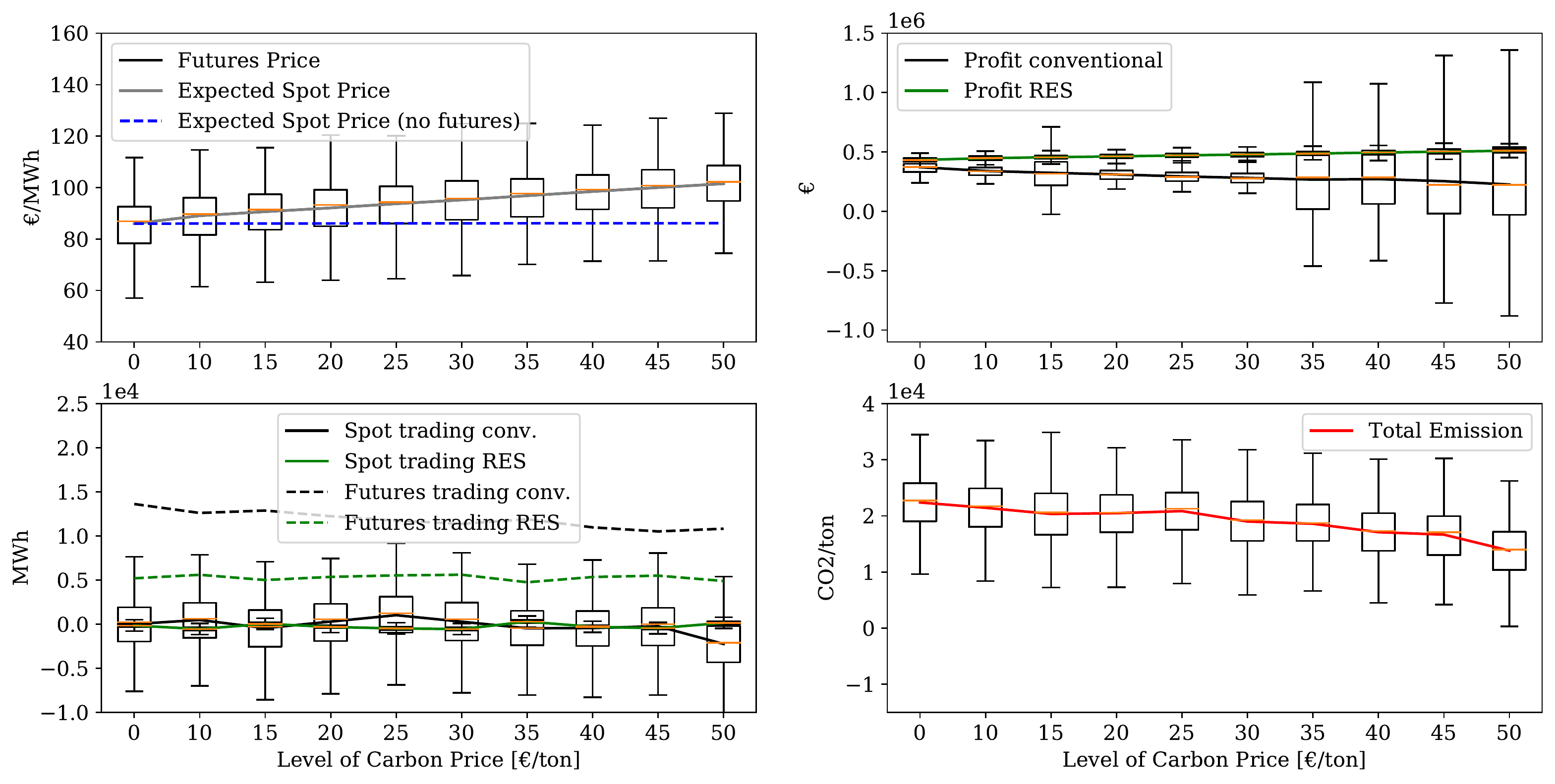}
        \caption{Risk neutral perfect competitive model with sensitivity on CO$_2$ price}
        \label{fig_s4}
    \end{figure}
\subsubsection{Levels of Competition and Policy Implication}
 RES penetration and CO$_2$ price increase affect electricity prices in both levels of competition. RES penetration and CO$_2$ price increase have an opposite impact on prices. RES penetration reduces prices in both stages/levels of competition. Whereas the surge in CO$_2$ price increases the overall price. In the Cournot competition, prices are relatively higher than the perfect competition. Both parameters reduce total emissions trading. With RES penetration, generators trade emissions only in the spot market with a clear diminishing trend. Likewise, total emissions trading decreases with CO$_2$ price increase.
 
  From the social welfare point of view, the auction-based allowance trading transfers part of generators' profit into social welfare, hence avoiding windfall profit. This helps to reduce GHG emissions with both policies. From the generators' point of view, RES generators penetrate with low-carbon and sustainable generation. These generators obtain higher profits as they do not have production and allowance permits related costs. With both policies, the conventional generators face challenges to stay in the market with competition. 
\begin{table}
\resizebox{\textwidth}{!}{%
\begin{tabular}{lllllll}
\hline
\multirow{3}{*}{\begin{tabular}[c]{@{}l@{}}Level of competition\\ with parameter\end{tabular}} &
  \multicolumn{2}{c}{$P^F$} &
  \multicolumn{2}{c}{$P_{\omega}^S$ the GM} &
  \multicolumn{2}{c}{$P_{\omega}^S$ the spot only} \\ \cline{2-7} 
                 & \multicolumn{6}{c}{Risk} \\ \cline{2-7} 
                 & Neutral & Averse  & Neutral & Averse    & Neutral     & Averse \\ \hline
Cournot RES& 103-126& 103-122 & 84-113& 83-107 & 86-113& 84-110\\
Competitive RES & 85-106& 87-108& 85-106 & 83-105 & 77-98& 76-97\\ \hline
Cournot CO$_2$& 108-120 & 103-116 & 91-105 & 88-103 & 99-100& 97-98\\
Competitive CO$_2$ & 85-101& 90-106& 85-101& 86-102 & 85-86& 86-87\\ \hline
\end{tabular}}
\caption{The minimum and the maximum futures price and expected spot prices,for the GM and spot only market, by competitions with RES penetration and CO$_2$ price increase.}
\label{tab_14}
\end{table}
    \subsection{Risk Averse Generators' Numerical Results}
This section discusses the risk averse generators simulation results where generators maximize the CVaR with the risk parameter $\phi=1$. Computationally, the risk averse model works with a large number of scenarios ($|\Omega| =320$). In this section, we present the risk averse generators' results based on sensitivities on the two parameters.
\subsubsection{Sensitivity on Renewable Penetration}\label{4.4.1}
    \begin{figure}
        \centering
        \includegraphics[width=\textwidth]{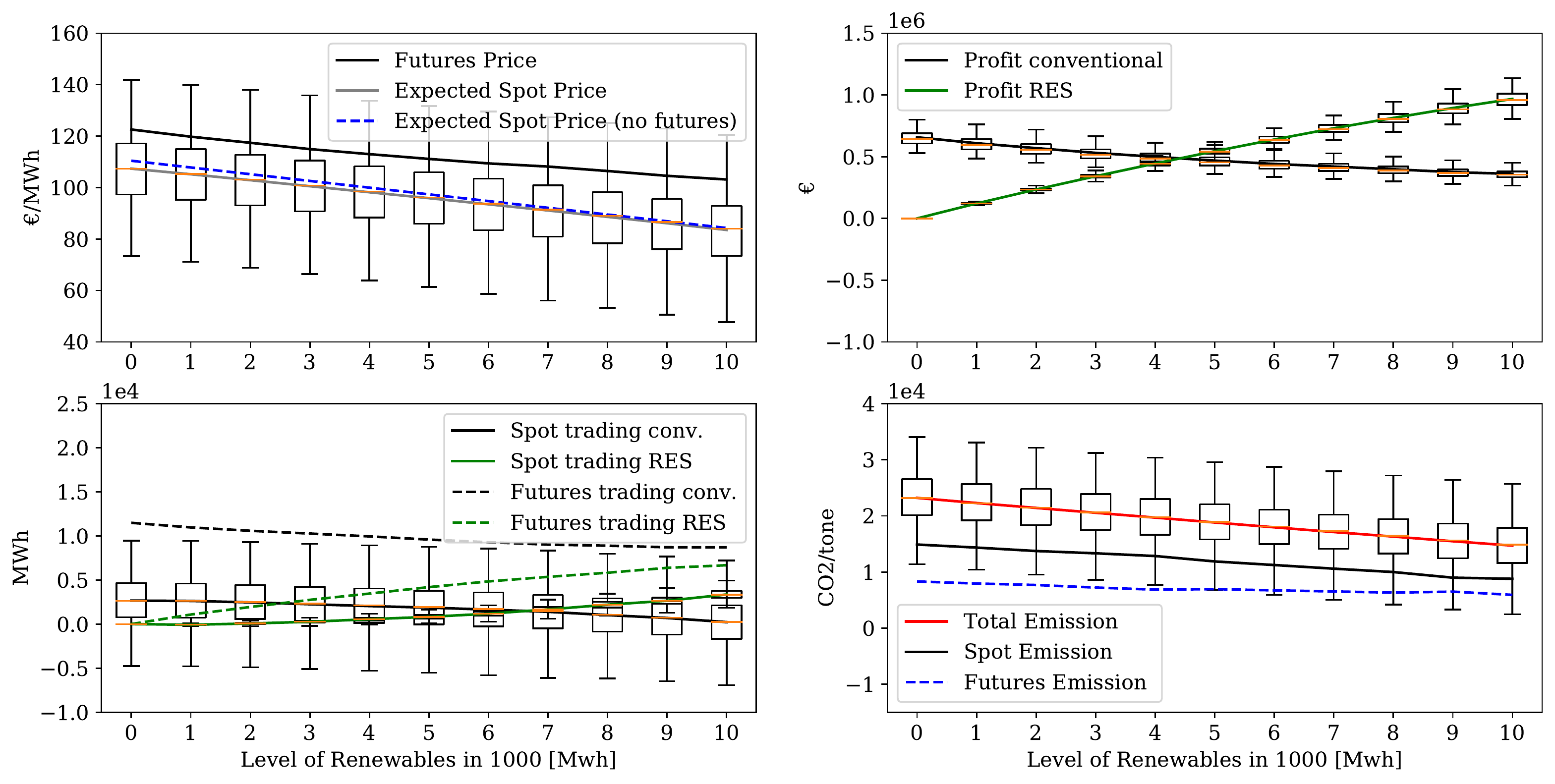}
        \caption{Risk Averse Cournot model simulation results with sensitivity analysis on RES penetration}
        \label{fig_s5}
    \end{figure}
        \begin{figure}
        \centering
        \includegraphics[width=\textwidth]{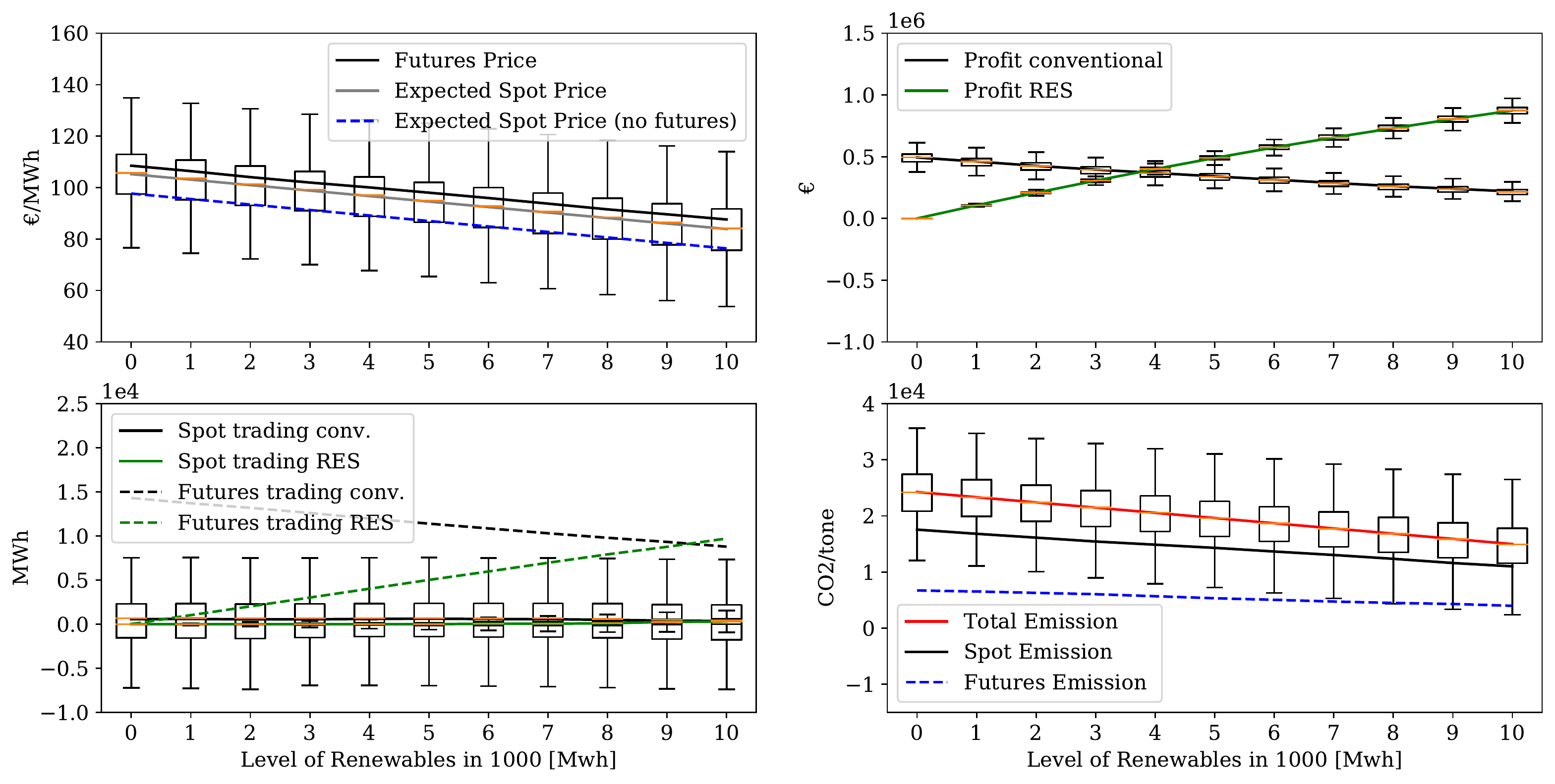}
        \caption{Risk Averse perfect competition model with sensitivity analysis on RES penetration}
        \label{fig_s6}
    \end{figure}
\cref{fig_s5} and \cref{fig_s6} illustrate the Cournot competition and perfect competition results for the RES penetration, respectively. In the Cournot competition with risk aversion, the expected spot and futures price behavior is similar to the risk neutral case. The expected spot prices and the futures price decrease from 107 to 83\euro/MWh, and from 126 to 103\euro/MWh, respectively. Unlike the risk neutral case with Cournot competition, conventional generators' futures trading slightly decreases. Spot trading is slightly affected by RES penetration especially when the level of RES penetration is greater than 7000MWh. Both spot and futures tradings with respect to RES generators increase. The loss in terms of quantities for conventional generators is compensated by the increase in the RES generators' quantities in futures and spot markets. We can then conclude that in the risk aversion with Cournot competition RES penetration is an effective policy, which is a counter-intuitive result of this work.

In the perfect competition, we can observe that the futures and expected spot prices are no more overlapped. $P^F $ and $P_{\omega}^S$ in the GM and $P_{\omega}^S$ in the spot only market are within the range of [87,108],[83,105] and [76,97], respectively (see \cref{tab_14}). For trading quantities, RES penetration increases the traded quantities for RES generators in the futures market and decreases futures trading for conventional generators. Trading quantities in the spot market remain unchanged for both sets of generators. Since the decrease in electricity generation for conventional generators is substituted by the RES, the consumers’ welfare is unaffected as large quantities are traded with relatively lower competition prices than the Cournot competition.
\subsubsection{Sensitivity on Carbon Price Increase}\label{4.4.2}
\begin{figure}
        \centering
        \includegraphics[width=\textwidth]{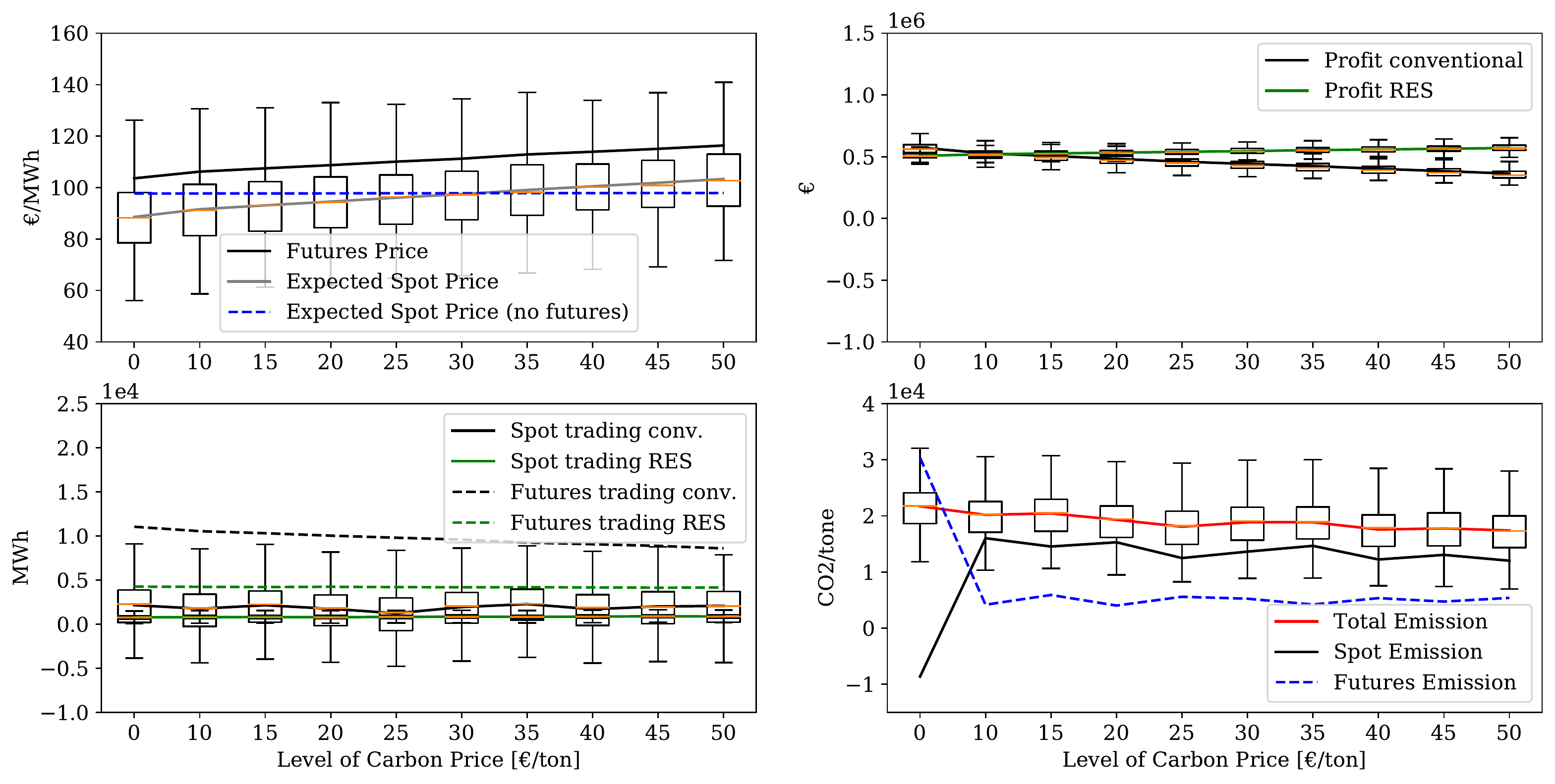}
        \caption{Risk Averse Cournot model with sensitivity on CO$_2$ price.}
        \label{fig_s7}
    \end{figure}

\begin{figure}
        \centering
        \includegraphics[width=\textwidth]{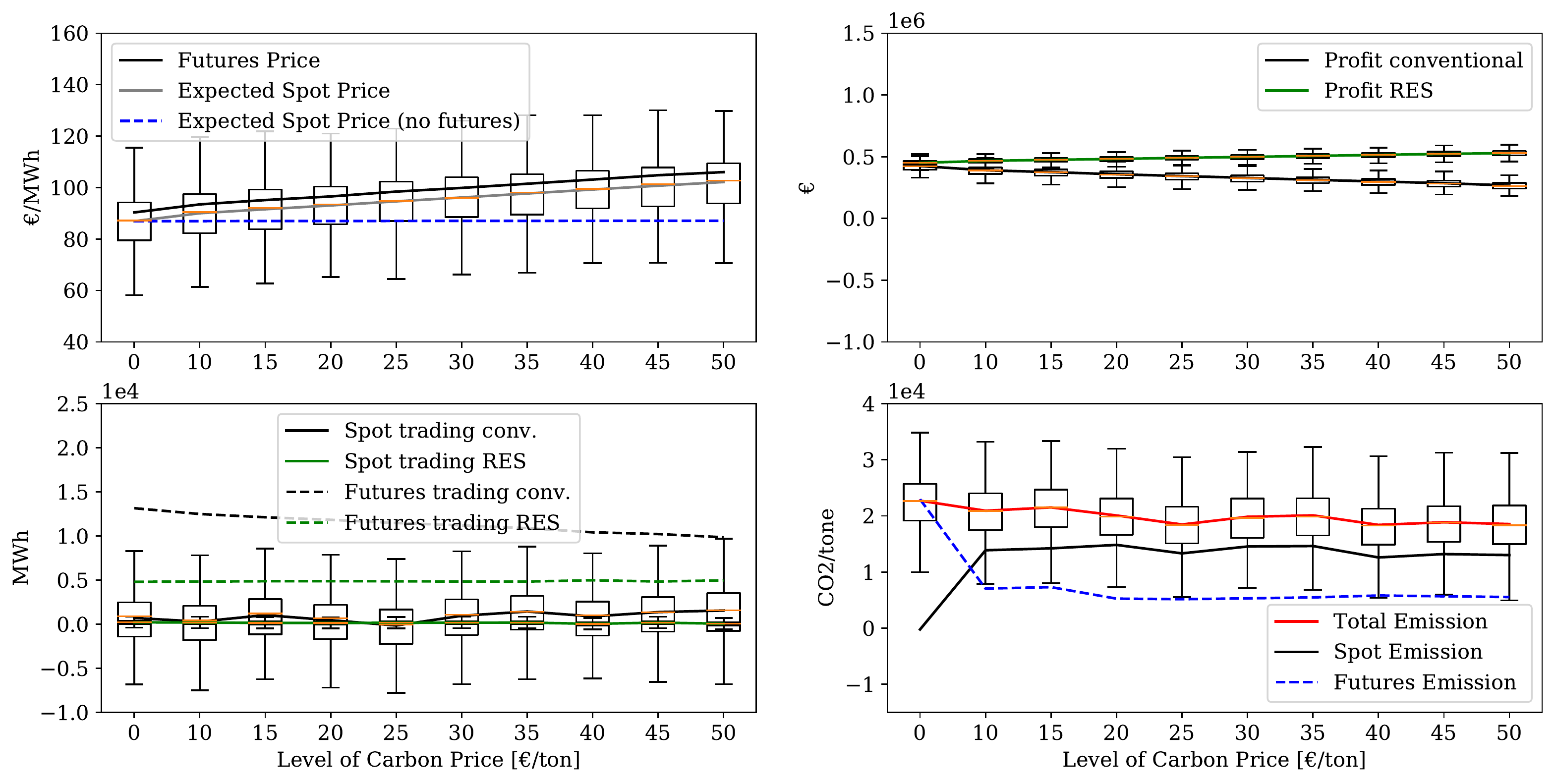}
        \caption{Risk Averse perfect competitive model with sensitivity on CO$_2$ price}
        \label{fig_s8}
    \end{figure}

\cref{fig_s7} and \cref{fig_s8} show the results of the increase in CO$_2$ price for Cournot and perfect competition levels, respectively. In the Cournot competition, the effect of an increase in the CO$_2$ price on profit is slightly different compared to the risk neutral case. Risk averse conventional generators' profit is decreased more than the risk neutral ones. Similar to the risk neutral case, futures conventional generators trading slightly decreases and spot market tradings are not significantly affected.

For the perfect competition, the expected spot and futures market prices are not overlapped. The higher CO$_2$ price is partially recovered by higher futures market electricity price that ranges from 90 to 106\euro/MWh. The increase in CO$_2$ price increases the overall prices in both stages of the market. This electricity price increase corresponds with the decrease in the futures trading. Spot trading for conventional generators is not affected by the increase in CO$_2$ price. However, the expected spot price slightly increases for CO$_2$ price increase. This is induced by the cost increase for conventional generators so that is a cost pass-through to the consumers. 

Differently from the risk neutral case in the Cournot competition (as examined above) conventional generators react to the higher CO$_2$ prices by slightly increasing their spot trading and reducing futures trading. This reduces the impact on the total profit. RES generators benefit from this policy where their profit increases with respect to CO$_2$ price increase, though no significant increase in their generation.
It is possible to conclude that the risk averse setting in both competitions CO$_2$ price increase is less effective policy than RES penetration.
\subsubsection{Levels of Competition and Policy Implication}   

\begin{table}
\resizebox{\textwidth}{!}{%
\begin{tabular}{lcccc}
\hline
\multicolumn{1}{c}{\multirow{3}{*}{Variables}} & \multicolumn{4}{c}{Parameters} \\ \cline{2-5} 
\multicolumn{1}{c}{}& \multicolumn{2}{c|}{RES Penetration ([5,000-10,000]MWh)} & \multicolumn{2}{c}{CO$_2$ price increase ([25-50]\euro/MWh)} \\ \cline{2-5}  &
  \begin{tabular}[c]{@{}c@{}}Cournot\\  Model\end{tabular} &
  \multicolumn{1}{c|}{\begin{tabular}[c]{@{}c@{}}Competitive\\  model\end{tabular}} &
  \begin{tabular}[c]{@{}c@{}}Cournot \\ Model\end{tabular} &
  \begin{tabular}[c]{@{}c@{}}Competitive \\ model\end{tabular} \\ \hline
$P_{\omega}^S$    & 92.70   & \multicolumn{1}{c|}{86.54}     & 94.47      & 95.96     \\
$P^F$    & 107.14 & \multicolumn{1}{c|}{92.74}     & 113.89     & 103.09     \\
total $\sum_{i}^{I}(\varepsilon_{i\omega}^S+\varepsilon_i^F)$   & \textbf{16,730.78} & \multicolumn{1}{c|}{\textbf{17,289.33}}   & \textbf{18,088.58} & \textbf{19,142.94} \\
$\sum_{i}^{I}q_i^F$   & 9,028.78 & \multicolumn{1}{c|}{10,069.13}   & 9,064.43     & 10,514.06   \\
$\sum_{i}^{I}q_{i\omega}^S$  & 1,138.59 & \multicolumn{1}{c|}{491.94}    & 1,988.28    & 1,218.53      \\
total $\sum_{i}^{I}(q_i^F+q_{i\omega}^S)$  & \textbf{10,167.37}& \multicolumn{1}{c|}{\textbf{10,561.07}}& \textbf{11,052.71}& \textbf{11,732.59}  \\
$\sum_{j}^{J}q_j^F$ & 5,542.88 & \multicolumn{1}{c|}{7,382.86}   &  4,157.07    & 4,866.1    \\
$\sum_{j}^{J}q_{j\omega}^S$  & 1,975.83 & \multicolumn{1}{c|}{115.64}     & 849.28     &  128.74       \\
total $\sum_{j}^{J}(q_j^F+q_{j\omega}^S)$  & \textbf{7,518.71}  & \multicolumn{1}{c|}{\textbf{7,498.50
}}     & \textbf{5,006.35}   & \textbf{4,994.84}      \\
$\sum_{i}^{I}\Pi_{i\omega}$  &410,644.03 & \multicolumn{1}{c|}{277,124.65} & 402,861.70 & 299,116.87  \\
$\sum_{j}^{J}\Pi_{j\omega}$   & 765,046.94 & \multicolumn{1}{c|}{688,389.88} &558,535.16 & 514,268.11   \\
$\sum_{i}^{I}CVaR_i$  & 317,690.49 & \multicolumn{1}{c|}{190,997.50}  & 305,364.36   & 207,757.98    \\
$\sum_{j}^{J}CVaR_j$  &688,191.82 & \multicolumn{1}{c|}{624,651.80} & 509,343.34    & 464,237.22    \\ \hline
\end{tabular}}
\caption{Expected values of the market outcomes with the CVaR formulation.}
\label{tab_15}
\end{table}
 \cref{tab_14} and \ref{tab_15} are useful to compare the Cournot and the perfect competitions based on the two sensitivities. With RES penetration, the expected spot electricity price is lower in the perfect competition that range from 83 to 105 \euro/MWh, and futures price varies from 87 to 108\euro/MWh. Conventional generators earn higher profit in Cournot competition with RES penetration. This higher profit is induced by the higher futures price ([103,122]\euro/MWh), and higher expected total tradings (11,480.01MWh). Similarly, with CO$_2$ price increase, the Cournot competition generates higher profit as a result of a higher $P^F$ ([103,116]\euro/MWh). RES generators also trade lower total quantities in the Cournot competition but manage to obtain a higher profit with respect to CO$_2$ price increase. Measured with profit, Cournot competition is better for both sets of generators. However, the perfect competition prevails lower electricity prices with respect to both policies. 

\cref{tab_15} summarizes the two policies' impact on market outcomes when the sensitivities are considered above the values of their respective average. Accordingly, RES penetration decreases total emissions as the conventional generation decreases. However, it increases RES generators' trading share in the market. Note that despite the relatively higher total tradings, both RES penetration and CO$_2$ price increase in the perfect competition reduces the quantities sold in the spot market for both sets of generators. In other words, higher total quantities are traded in perfect competition than the Cournot competition with respect to both parameters. Electricity prices are lower since there is no cost pass through. More interestingly, RES generators expected profit is higher than their conventional counterparts, which shows both policies are effective in achieving emission reduction targets and green economic path. However, in the grand scheme of things, RES penetration seems to be more effective than CO$_2$ price increase since it results in lower emissions, lower electricity prices, and higher trading quantities.
\section{Summary and Conclusion}\label{S_5}
This paper proposes a model based on game theory to assess the effects of CO$_2$ auctioning with futures market contract designs in a stochastic programming approach, with high renewable penetration in oligopolistic electricity markets. The game is based on electricity derivatives and EUA allowances, applying for different levels of competition (Cournot and perfect competition) in a two-stage stochastic model.
    
We examine risk neutral/averse stochastic and oligopolistic generators, with the option of primarily trading their production and EUA allowances in a futures market (stage-one) and later in a spot market (stage-two) anticipating the inherent uncertainties. The sources of uncertainties considered are electricity, EUA allowance prices, generation cost for conventional generators, and level of nondispatchable RES generation.

A coherent risk measure (CVaR) is used in the model to characterize the behavior of risk averse generators. The futures contracts are introduced for both electricity and allowances (to trade surplus/shortage of allowances in the auction market) to hedge the risk of both prices and RES generation capacity in the equilibrium market model. 
    
Analytically, we have closed-form solutions in the second stage, where we start to derive and move backward to characterize the first stage variables in terms of the second stage equilibrium market outcomes. The global equilibrium of the market is computed from the joint solution of all the generators' profit maximization problems through maximizing the sum of expected profits for risk neutral generators and maximizing the CVaR for risk averse generators. This is done by solving an equivalent system of optimality (KKT) conditions. 
    
The analytical results are checked with a wide computational test to analyze different market configurations, Cournot, and perfect competitive generators together with risk neutrality, or risk averse strategies. 
    
We conclude that RES penetration is an effective and economically efficient parameter in plummeting GHG emissions. The increase of CO$_2$ emission prices, on the other hand, encourages RES generators by strategically penalizing CO$_2$ emitters from the market so that it decreases futures and spot emissions. The result corroborates with other findings such as RES deployment and an increase of CO$_2$ prices lead to higher energy-intensive sectors to leave the industry and lower energy-intensive sectors to be attracted to enter the economy.
Future research may include to study the financial and environmental impact of emerging trends in power systems, such as distributed generation, large-scale electricity storage, demand response programs, etc.
\color{black}
	\bibliographystyle{elsarticle-num-names}
	\newpage
	\bibliography{Reference-WP}
\end{document}